\RequirePackage{etex}
\documentclass[fleqn,a4paper]{IEEEtran}
\IEEEoverridecommandlockouts
\usepackage{subfig}
\usepackage[table,dvipsnames]{xcolor}
\usepackage{amssymb,csquotes,enumitem,amsthm,cite,booktabs,tikz,mathtools,multirow,stfloats}
\usepackage{algorithmic,graphicx,textcomp,url,doi,hyperref,cleveref,float,placeins,pifont,tabularx}
\usepackage{tikz-network}
\usepackage{calrsfs}
\DeclareMathAlphabet{\pazocal}{OMS}{zplm}{m}{n}
\usepackage[T1]{fontenc}
\hypersetup{
    colorlinks,
    linkcolor={blue!50!black},
    citecolor={blue!50!black},
    urlcolor={blue!80!black}
}
\usetikzlibrary{arrows,positioning,shapes,calc}
\renewcommand{\(}{\left(}
\renewcommand{\)}{\right)}
\newcommand{\qt}{\enquote}
\newcommand{\nscale}{\bar{n}}
\newcommand{\N}{\mathbb N}
\newcommand{\NN}{\pazocal N}
\newcommand{\R}{\mathbb R}
\newcommand{\RR}{\pazocal R}

\renewcommand{\O}{\pazocal O}

\newcommand{\F}{\pazocal F}
\newcommand{\X}{\pazocal{X}}
\newcommand{\A}{\pazocal A}
\renewcommand{\F}{\pazocal F}
\renewcommand{\P}{\mathbb P}
\newcommand{\state}{X}
\newcommand{\rstate}{X}
\newcommand{\states}{\X}
\renewcommand{\S}{\mathcal{S}}
\renewcommand{\b}{\bfseries}

\newcommand{\tabitem}{\llap{\textbullet}}
\renewcommand{\pod}{\operatorname{PoD}}
\graphicspath{{figures/}{./}}
\DeclareMathOperator*{\argmax}{arg\,max}

\definecolor{LightCyan}{rgb}{0.88,1,1}
\definecolor{LightCyan2}{rgb}{0.1,1,1}
\definecolor{ppcolor}{rgb}{0.212,0.506,0.584}
\definecolor{pcolor}{rgb}{0.255,0.596,0.686}
\definecolor{rcolor}{rgb}{0.294,0.675,0.776}
\definecolor{ecolor}{rgb}{0.569,0.765,0.835}
\definecolor{scolor}{rgb}{0.733,0.843,0.890}
\definecolor{ocolor}{rgb}{0.783,0.893,0.940}
\definecolor{oocolor}{rgb}{0.783,0.893,0.940}
\newcommand{\rowdark}{\rowcolor{scolor}}
\newcommand{\bottomline}{\noalign{\global\arrayrulewidth=0.5mm}\arrayrulecolor{scolor}\hline}
\newcommand{\rowlight}{\rowcolor{oocolor}}
\newcolumntype{L}{>{\hspace*{-0.6\tabcolsep}}l}
\newcolumntype{R}{r<{\hspace*{-0.6\tabcolsep}}}
\newcolumntype{C}{c<{\hspace*{-0.6\tabcolsep}}}
\theoremstyle{definition} 
\newtheorem{definition}{Definition}
\newtheorem{example}{Example}

\newcommand{\PLUS}{+}

\newcommand{\MINUS}{--}
 
\newcommand{\trot}[1]{\multicolumn{1}{l}{\rlap{\rotatebox{60}{#1}~}}} 
\newcommand{\spemph}[1]{\textcolor{ppcolor}{\emph{#1}}}

\begin{document}

\title{PREStO: A Systematic Framework for Blockchain Consensus Protocols\thanks{This work was supported in part by the National Research Foundation (NRF), Prime Minister's Office, Singapore, under its National Cybersecurity R\&D Programme (Award No. NRF2016NCR-NCR002-028) and administered by the National Cybersecurity R\&D Directorate. Georgios Piliouras acknowledges SUTD grant SRG ESD 2015 097, MOE AcRF Tier 2 Grant 2016-T2-1-170 and NRF 2018 Fellowship NRF-NRFF2018-07.}}

\author{\IEEEauthorblockN{Stefanos Leonardos, Dani\"el Reijsbergen, and Georgios Piliouras} \\
\IEEEauthorblockA{Singapore University of Technology and Design}}

\maketitle
\begin{abstract} 
The rapid evolution of blockchain technology has brought together stakeholders from fundamentally different backgrounds. The result is a diverse ecosystem, as exemplified by the development of a wide range of different blockchain protocols. This raises questions for decision and policy makers: How do different protocols compare? What are their trade-offs? Existing efforts to survey the area reveal a fragmented terminology and the lack of a unified framework to reason about the properties of blockchain protocols. \par
In this paper, we work towards bridging this gap. We present a five-dimensional design space with a modular structure in which protocols can be compared and understood. Based on these five axes -- \emph{Optimality, Stability, Efficiency, Robustness and Persistence} -- we organize the properties of existing protocols in subcategories of increasing granularity. The result is a dynamic scheme -- termed the \emph{PREStO framework} -- which aids the interaction between stakeholders of different backgrounds, including managers and investors, and which enables systematic reasoning about blockchain protocols. We illustrate its value by comparing existing protocols and identifying research challenges, hence making a first step towards understanding the blockchain ecosystem through a more comprehensive lens.
\end{abstract}



\section{Introduction}
\label{sec:introduction}
In the seminal Bitcoin paper \cite{Na08}, the pseudonymous Satoshi Nakamoto pioneered the use of \emph{blockchains} as a secure way of maintaining a ledger of currency transfers in a trustless peer-to-peer network.
In the ten years since, blockchains have grown \cite{mergers} to underpin a \$100 billion cryptocurrency market \cite{coinmarketcap}. Meanwhile, their applicability is increasingly understood in a broad range of other contexts \cite{Ca19}, e.g., the Internet of Things \cite{ferrag2018blockchain}, supply chain management \cite{kshetri20181}, healthcare \cite{mettler2016blockchain}, etc.
This rapid growth has induced a considerable number of established market parties to invest in the sector \cite{nasdaq,fidelity}, or even develop their own platforms. Noteworthy examples of the latter include \emph{Quorum} \cite{quorum}, which is developed by JPMorgan Chase, and the \emph{HyperLedger} umbrella project \cite{Cac16}, hosted by the Linux Foundation and supported by, inter alia, IBM and Intel. Applications of Quorum include JPMorgan's internal digital currency \cite{jpmcoin} and the Interbank Information Network \cite{iinquorum,jpmiin}, a platform for cross-border money transfers. Applications of HyperLedger include a project by the US retailer Walmart to track the movement of vegetables \cite{hlwalmart,walmart}. IBM by itself had 1500 employees working on 500 blockchain-related projects in September 2018 \cite{ibm}. Meanwhile, new multipurpose blockchain platforms developed by startups continue to emerge, e.g., \emph{Ethereum} \cite{Bu14online}, \emph{Cardano} \cite{cardano,kiayias2017ouroboros}, \emph{Algorand} \cite{algorandfunds,Gi17}, and \emph{Zilliqa} \cite{zilliqa,zilliqawp}.\par
This proliferation of blockchain technologies and applications has brought together stakeholders with fundamentally different degrees of technical expertise. So far, the discourse between these groups has been marked by the use of sometimes incongruous terminology, and the lack of a unified communication framework \cite{Pri18online}. This hampers the ability of managers and investors to make business decisions, and of newly proposed protocols to be compared and understood. Particularly affected are one of the most fundamental technical aspects of blockchain platforms: the \emph{consensus protocols}.

\begin{figure}[!htb]
\centering
\begin{tikzpicture} 
\draw[black, fill = pcolor] (0,0) arc (-90:270:3cm and 2.5cm);
\draw[black, fill = rcolor] (0,0) arc (-90:270:2.5cm and 2cm);
\draw[black, fill = ecolor] (0,0) arc (-90:270:2cm and 1.5cm);
\draw[black, fill = scolor] (0,0) arc (-90:270:1.5cm and 1cm);
\draw[black, fill = ocolor] (0,0) arc (-90:270:1cm and 0.5cm);

\node at (0,0.5) {\textbf{O}ptimality};
\node at (0,1.5) {\textbf{S}tability};
\node at (0,2.5) {\textbf{E}fficiency};
\node at (0,3.5) {\textbf{R}obustness};
\node at (0,4.5) {\textbf{P}ersistence};

\end{tikzpicture}
\caption{The PREStO framework as a nesting doll of goals.}
\label{fig:spore}
\end{figure}
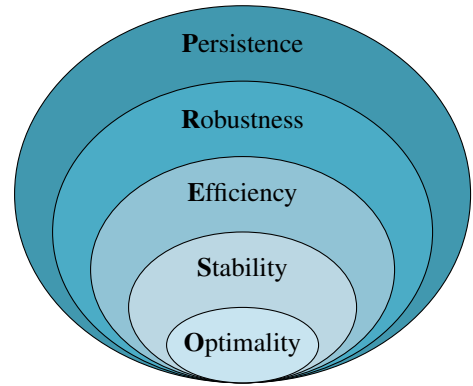

Consensus protocols fulfill, in a decentralized setting, the role that a single authority has in a centralized database or ledger. It is the mechanism to reach agreement among self-interested peers, and for making consistent decisions out of mutually exclusive alternatives. The choice of consensus protocol has a major impact on a platform's performance, including its security and throughput, and is therefore important for anyone who is involved in blockchain development \cite{Del19online}, particularly executives. This can be challenging if the differences between the alternatives are not well-understood.
\subsection*{Statement of Contribution and Managerial Relevance}
In this paper, we address these difficulties by developing an accessible, yet technical and comprehensive framework to improve the communication between the diverse participants of the blockchain ecosystem. We assume only a basic understanding of mathematics and the high-level idea behind blockchains, and introduce technical terms related to blockchains and cryptocurrencies from the bottom up.\par
Our main contribution is the \emph{PREStO framework}~\cite{Pz17,chia2018rethinking}, which is a dynamic tool to identify and classify properties of blockchain protocols. It is an acronym (in reverse order) of its five main axes: Optimality, Stability, Efficiency, Robustness and Persistence, cf. \Cref{fig:spore}. 
In \Cref{tab:questions}, we capture the essence of each category in a single question.
\begin{table}[!htb]
\vspace{0cm}\centering
\setlength{\tabcolsep}{6pt}
\renewcommand{\arraystretch}{1.2}
\begin{tabularx}{\linewidth}{LXC}
\rowdark
\b Dimension & \b Description & \b Section\\\\[-0.3cm]
\rowlight
\b Optimality & Does the protocol maximize the quality of its core outcomes under normal circumstances? & \ref{sec:optimality} \\
\b Stability & Is the designed protocol an equilibrium? & \ref{sec:stability} \\
\rowlight
\b Efficiency & How does the protocol utilize its different resources, e.g., time, space, energy, network bandwidth? & \ref{sec:efficiency}  \\
\b Robustness & Does the protocol's performance withstand perturbations to its parameters? & \ref{sec:robustness} \\
\rowlight
\b Persistence & If the protocol is forced out of equilibrium, does it recover? & \ref{sec:persistence} \\\\[-0.3cm]
\bottomline
\end{tabularx}\\[0.2cm]
\caption{The five axes of PREStO and their main purpose.}\vspace{-0.2cm}
\label{tab:questions}
\end{table}

PREStO's modular structure sets it apart from related efforts and enhances its value for managers. Initially, the five categories can be seen as a nesting doll of design goals, where each category considers a wider range of desirable properties than the previous, cf. \Cref{fig:spore}. We start at the very basic -- i.e., optimal performance under ideal conditions -- and gradually build up to the more advanced -- e.g., recovery mechanisms to survive in the long run.
Subsequently, the axes are organized into subcategories of increasing granularity, and PREStO develops into a dynamic tool to identify and group together challenges and research opportunities for the various blockchain protocols, cf. \Cref{tab:challenges}. We demonstrate its practical use via two running use cases, Bitcoin and Quorum, and conclude with a schematic illustration of the resulting classification in \Cref{fig:presto_detailed}.
Furthermore, we extensively draw from the existing literature to motivate our framework.

\subsection*{A Growing Ecosystem}
The consensus protocol introduced by the first blockchain platform -- Bitcoin -- is commonly called \emph{Nakamoto consensus} \cite{St18}. It was designed to work in a \emph{permissionless} setting, i.e., a setting in which any node in the network is allowed to add data to the blockchain. To prevent network overflow, nodes who seek to extend the blockchain must spend computational effort through a process called \emph{mining}. In the presence of competing chains, honest nodes accept the chain with the most effort spent on creating it.
Together, these rules ensure that if more than 50\% of the computational power is in the hands of honest parties, then their chain will grow faster than all others.
Nodes are compensated for the spent computational power through \emph{rewards} in the form of \emph{tokens} logged on the blockchain.
Variations of Nakamoto consensus are currently implemented in over 600 cryptocurrencies \cite{Zh19,coinmarketcap}, including the Ethereum platform and various Bitcoin spin-offs.\par
In recent years, Nakamoto consensus has increasingly drawn criticism for its \emph{low transaction throughput} and \emph{high energy consumption}. A single Bitcoin transaction costs more energy than 100\,000 Visa transactions, and the Bitcoin network as a whole consumes as much energy as a medium-sized country \cite{Di18online}.
Furthermore, it is \emph{insecure} in the sense that smaller platforms are vulnerable to attackers who seize a majority of the computational power, as witnessed by the recent 51\% attacks on Ethereum Classic \cite{Je19online} and Bitcoin Gold \cite{Ro18online}. 
Finally, research \cite{Ey14,Ge16,Sa17,Zh19} has shown that Nakamoto consensus can be \emph{incentive-incompatible}, i.e., participants can increase their rewards by deviating from the protocol.
To address these weaknesses, a multitude of new consensus protocols have been proposed that more closely follow traditional theory on \emph{permissioned} (i.e., not open) networks. In particular, many approaches use variations of \emph{Byzantine fault tolerant} (BFT) protocols \cite{lamport1982byzantine} or other classical consensus protocols such as Paxos \cite{lamport1998part} and Raft \cite{ongaro2014search}. 
Such approaches can achieve gains in efficiency and security at the cost of centralization. However, a precise description of this trade-off is complicated due to the differences between BFT protocol implementations, and the lack of alignment between the terminology used by different parties. This motivates the need for a formal framework to describe and compare different consensus protocols.

\subsection*{Related Work}
The necessity of developing a unified communication framework for the blockchain ecosystem was already acknowledged in \cite{Za18}. Accordingly, a brief outline of the PREStO framework was first introduced in \cite{chia2018rethinking}. While the five main axes remain the same, their organization into subcategories is first deployed in the present paper. \par

The rapid growth of the blockchain-related literature has also stimulated other projects that survey the area from different perspectives. Focusing exclusively on the Bitcoin blockchain, \cite{Co18} provide a systematic review of Bitcoin's underlying features, particularly its security and privacy-related threats and vulnerabilities, and discuss directions for future research. Their analysis extends initial analyses of the backbone protocols of the main cryptocurrencies \cite{Ga15,Ga17,Pa17a, Leo18}. In \cite{St18}, further insight is provided into the development and functionality of the Bitcoin blockchain, in addition to a non-exhaustive, yet interesting timeline of papers related to the analysis of Nakamoto consensus. \par

In a spirit closer to the present study, \cite{Wa18} acknowledge the lack of a comprehensive literature review on the various layers of blockchain technology, and provide a rigorous vision on the organization of blockchain networks. Their work extends to all aspects of the relevant technology and provides a central reference for future work. They define four layers for any blockchain system, from top to bottom: (1) the application layer, (2) the virtual machine layer, (3) the consensus layer, and (4) the network layer. In the present study, we focus on the third (i.e., consensus) layer. That is, application-layer properties, virtual-machine-layer properties (e.g., secure smart contract languages such as Scilla \cite{sergey2018scilla}), and network-layer properties (e.g., vulnerability to eclipse \cite{He15}, BGP hijacking \cite{Ap16}, or DoS \cite{Jo14} attacks) are treated only if and when they affect the consensus layer.

 The difficulty to conceptualize the dramatically evolving design landscape of blockchains is further supported by \cite{Ba17}. Similar to the present work, they focus on the consensus layer and discuss the various themes and key approaches that are exhibited by current blockchains. They systematize distinctive features and technical properties of existing consensus protocols and provide thorough comparisons, open questions, and directions for future research. Despite the common perspectives, our approach distinguishes itself from \cite{Ba17} due to its mathematical framework that allows for a description of properties from the ground up.\par

Using a practice-oriented focus, \cite{Din17} develop BLOCKBENCH, a promising and publicly available software program that is designed to test and compare the performance of blockchain protocols. It applies to private blockchains and its findings are mainly associated with properties in the categories of Optimality and Efficiency of the PREStO framework, cf. \Cref{sec:optimality,sec:efficiency}. The paper features use cases of the Ethereum, Parity and Hyperledger blockchains and concludes that these systems are still far from large-scale adoption. Finally, a non-exhaustive list of related surveys with focal points ranging from smart contract execution to general blockchain applications and research perspectives includes \cite{Bo15,Cr16,Yl16,At17,Ca17,Ka17,Zh18,Di18,Th18,Gar18,Ca19,azouvi2019sok,shahaab2019applicability}. 

\subsection*{Outline}
The current paper is structured as follows. We begin by defining an abstract, high-level model of a blockchain consensus protocol in \Cref{sec:model}. In \Cref{sec:optimality,sec:stability,sec:efficiency,sec:robustness,sec:persistence}, we describe the main PREStO axes and define their subgategories. We summarize related issues and open questions related to each category in \Cref{sec:evaluation}. \Cref{sec:conclusions} concludes our analysis with general comments and directions for future work. 


\section{A Mathematical Model for Blockchain Consensus}
\label{sec:model}
\begin{figure*}[!htb]\vspace{-0.3cm}
\centering
\subfloat[][The peer-to-peer network]{\includegraphics[width=0.49\textwidth]{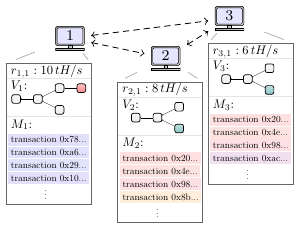}\label{fig:networkfigure}}\hspace{0.25cm}
\subfloat[][The blockchain DAG]{\includegraphics[width=0.49\textwidth]{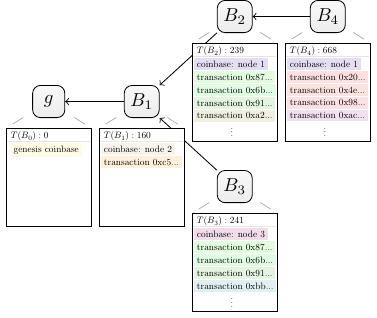}\label{fig:blockchainfigure}}
\caption{The state of a Bitcoin-like blockchain network 700 seconds after genesis ($t^*=700$): (a) the peer-to-peer network, including for each node $i$ their view $V_n^t$ (with the head $H^t_n$ colored), resources $r^t_n$, and memory pool $M_n^t$ (b) the blockchain itself, given as a graph $(V_t,E)$ where $V_t = \cup_n V_n^t $ and $(B,B') \in E$ iff $B' = P(B)$.}\vspace{-0.3cm}
\label{fig:networkandblockchainfigure}
\end{figure*}

To rigorously define the properties covered by the PREStO framework, we require a mathematical model to serve as a basis. In this section, we will present this model by describing the data on the blockchain (in \Cref{sec:blockdata}), the participating nodes as individuals (in \Cref{sec:nodes}), the nodes as a network (in \Cref{sec:network}), and the rewards and strategies (in \Cref{sec:incentives}). In all cases we denote the consensus protocol by $\Pi$.
To account for the wide variety of existing protocols and the diversity of their technical features, we keep our model as general as possible. However, we illustrate our presentation  using two running examples from the permissionless and permissioned settings, respectively. The first is Bitcoin \cite{Na08}, which is both the first and the (to our knowledge) simplest implementation of a permissionless protocol. The second is Quorum \cite{quorum}, which uses the Istanbul-BFT consensus protocol  \cite{istanbulbft} -- we chose this protocol in favor of the BFT features in HyperLedger because the latter were not well-documented at the time of writing, and in favor of other BFT-based blockchain protocols such as Tendermint because of Quorum's business focus. We also visualize the core concepts using \Cref{fig:networkandblockchainfigure} and \Cref{tab:evolution}, which display the state of a Bitcoin-like blockchain network at a given time $t^*$, and this network's evolution until and shortly after $t^*$, respectively. A list of the symbols introduced in this section is given in Table~\ref{tab:symbols}.

\begin{table}[!t]
\centering
\vspace{0.27cm}
\arrayrulecolor{pcolor}
\begin{tabular}{ll} 
\toprule
Symbol & Meaning \\ \midrule
$B \in \N$ & A block \\
$T(B)$ & Timestamp of block $B$ \\
$P(B)$ & Predecessor of block $B$ \\
$W(B)$ & Proof-of-work difficulty of block $B$ \\
$g$ & Genesis block \\
$C(B)$ & Chain between block $B$ and $g$ \\
$n, \NN$ & Node; set of active nodes \\
$i, \RR$ & Resource; set of protocol resources \\
$r^t_{n,i} \in \R^{+}$ & Amount of resource $i$ owned by node $n$ at time $t$\\
$R^t_{i} \in \R^{+}$ & Total amount of resource $i$ in the network at time $t$\\
$p^t_{n,i} \in [0,1]$ & Fraction of resource $i$ owned by node $n$ at time $t$ \\
$V^t_n$ & View of node $n$ at time $t$ \\
$M^t_n$ & Memory pool of node $n$ at time $t$ \\
$H^t_n$ & Head of node $n$ at time $t$ \\
$f$ & Fork-choice rule \\
$x, \X$ & State; state space \\
$s, S, \S$ & Strategy; strategy profile; space of all strategy profiles\\ 
$d$ & Default or \emph{follow-the-protocol} strategy \\
$v_n(S,x)$ & Long-term utility of node $n$ given strategy profile $S$ \\
& and initial state $x$\\
$\F_{\Pi}$ & Feature/property of protocol $\Pi$ \\
${U}_{\Pi}$ & Performance measure of protocol $\Pi$ \\
\bottomrule
\end{tabular}
\caption{List of symbols.}\vspace{-0.3cm}
\label{tab:symbols}
\end{table}

\subsection{Blockchains as Data Structures} \label{sec:blockdata}
At its core, a \spemph{blockchain} is a data structure that contains a sequence of elementary database operations called \spemph{transactions}. The semantics of the transactions depend on the platform \cite{Sou18} (e.g., they can represent token transfers, smart contract calls, sensor readings in an IoT context, etc).
The transactions are grouped into \spemph{blocks}. Each block not only contains transactions, but also a \spemph{header} that contains summary information about the block. The header of a block typically contains a reference to the block's transactions,\footnote{For example, via a Merkle tree root \cite{merkle1987digital}.} a \spemph{timestamp}, a reference to the \spemph{previous block}, and some additional platform-specific data cf.\ \cite{Br18}. We assume that each block can be identified by a unique integer in $\N$ (e.g., via the hash of its header).

In our model, we represent the block's timestamp as a function $T:\N\to \R^+$, i.e., if a block $B$ is created at time $t$ then $t=T\(B\)$.\footnote{In practice, the block creator has considerable freedom in choosing the timestamp \cite{szalachowski2018towards}.}
The previous block of any block $B$ is also represented via a function $P : \N \rightarrow \N$, i.e., $B$ points to $P(B)$ as its previous block.
For brevity, we write $P^2\(B\) = P\(P\(B\)\)$, $P^3\(B\) = P\(P\(P\(B\)\)\)$, etc. The first block $g$ is called the \spemph{genesis} or the \emph{genesis block}, and is the only block to not have a previous block, i.e., $P\(g\) = \emptyset$.
Using $P$, we can construct for each block $B$ a \spemph{chain} of blocks $C(B)$ to the genesis, i.e., $C\(B\) = \(B, P\(B\), P^2\(B\), \ldots, g\)$.
The chain is \spemph{cryptographically secure}, i.e., the relationship between a block $B$ and its predecessor $P(B)$ is given by encapsulating \emph{all} information in $P(B)$ in $B$ via a cryptographic \emph{hash function}. Essentially, if even a single bit of data in $P(B)$ is changed, then executing the hash function on its header will produce a different hash, and the relationship between $B$ and $P(B)$ is broken.

At any point in time, the complete structure of the blocks created so far can be represented using a \spemph{Directed Acyclic Graph (DAG)}, where the blocks are the vertices and a directed edge between blocks $B$ and $B'$ exists if $B' = P(B)$, see Figure~\ref{fig:blockchainfigure} for an example. In this paper, we only consider \spemph{chain-based} protocols, i.e., protocols for which the validity and the output of a transaction within any block $B$ depends only on $C(B)$. That is, we do not include protocols such as Avalanche \cite{rocket2018snowflake} and IOTA's Tangle, even though some of the definitions in the PREStO framework may still be applicable.

\setcounter{example}{0}
\begin{example}[Bitcoin] In Bitcoin, transactions represent token transfers between users. Transactions also include \spemph{fees} that are paid to block creators. Among the block header fields, the \spemph{proof-of-work}, $W: \N \rightarrow \R^+$ (as given via the difficulty and the nonce) is also relevant.
\end{example}

\begin{example}[Quorum] Quorum is based on Ethereum, and hence the transactions not only represent token transfers, but also smart contract calls and creations. The protocol messages of Instanbul-BFT, e.g., \emph{prepare} and \emph{commit} messages, are also seen as transactions, even though only the commit messages are included in the block (yet they are not referred to in the header).
\end{example}

\subsection{Nodes and Resources} \label{sec:nodes}

The blockchain protocol is operated by a set $\N$ of agents called \spemph{nodes}, which are identified by their index $n \in \N$. 
To participate, each node $n$ provides each of $m$ distinct \spemph{resources}. The amount of resource $i \in \RR:=\{1,\dots,m\}$ contributed by node $n$ is denoted by $r_{n,i} \in \mathbb R_+ $. 
Since resources change over time, we will write $\vec{r}^{\,t}_n=\(r^t_{n,1},\dots, r^t_{i,m}\)$ for the vector of resources of node $n$ at time point $t>0$. Accordingly, let $\vec{R}^t = ({R}_{1}^t, \dots, {R}_{m}^t):=$\mbox{$\sum_{n\in\N} \vec{r}_n^{\,t}$} be the vector of total protocol resources at timepoint $t>0$. We similarly define $p_{n,i}^{t}:= r_{n,i}^{\,t} / R^t_{i}$ as the fraction of resource $i$ owned by node $n$ at time point $t>0$. We say that a group of nodes $N \subset \N$ is \spemph{$\alpha$-strong} with $\alpha \in [0, 1]$ in terms of resource $i \in \RR$ if
$$
\sum_{n \in N} p_{n,i} \geq \alpha.
$$
The nation of being $\alpha$-strong can be generalized to multiple resources, but we omit this for the sake of brevity. We also omit the time superscript whenever it is unnecessary (like in the previous equation). Similarly, when only a single resource is critical to the consensus mechanism, we omit the subscript $i$ entirely and write $p_n$. If $r_{n,i}^t>0$ for all $i \in \{1,\dots,m\}$, then we will say that node $n$ is \spemph{active} at timepoint $t$.  
Nodes communicate with other nodes via a software application called a \spemph{client}.

\setcounter{example}{0}
\begin{example}[Bitcoin] In Bitcoin, we can identify three major \qt{types} of nodes: 
\begin{enumerate}[leftmargin=*,label=\arabic*.]
\item \emph{mining nodes}, who create new blocks by solving computational \qt{puzzles}, 
\item \emph{full nodes}, who verify blocks before accepting them (i.e., check whether all the included transactions are valid), and 
\item \emph{light nodes}, who only verify block headers, and are only interested in checking the inclusion of individual transactions via Simplified Payment Verification (SPV).
\end{enumerate}
The foremost resource $r_{n,1}$ of node $n$ is \emph{processing power}: typically a mining node will need a great amount of it (e.g., an ASIC rig), a full node a moderate amount (e.g., a high-end PC), and a light node very little (e.g., a smartphone).
The model for Bitcoin can further be extended by including \emph{bandwidth} as a separate resource $r_{n,2}$.
\end{example}

\begin{example}[Quorum] In a permissioned blockchain, there can still be full nodes and light nodes, but mining nodes are unnecessary and processing power is less important (yet still required to, e.g., verify signatures). The main resource is access to the private keys that allow for the creation of blocks, i.e., \emph{authority}. We model this in the following way: we assume that there are $k$ private keys for block creation. Then for all $n \in \N$ and $i \in \{1,\ldots,k\}$, it holds that $r_{n,i} = 1$ if node $n$ controls key $i$, and $r_{n,i} = 0$ otherwise. Furthermore, $r_{n,k+1}$ denotes the processing power of node $n$ and $r_{n,k+2}$ the bandwidth. Note that it is possible for more than one nodes to have access to the same private key, e.g., after a hack. 
\end{example}

\begin{table*}[!htb]
\makebox[\textwidth][c]{\includegraphics[width=0.95\textwidth]{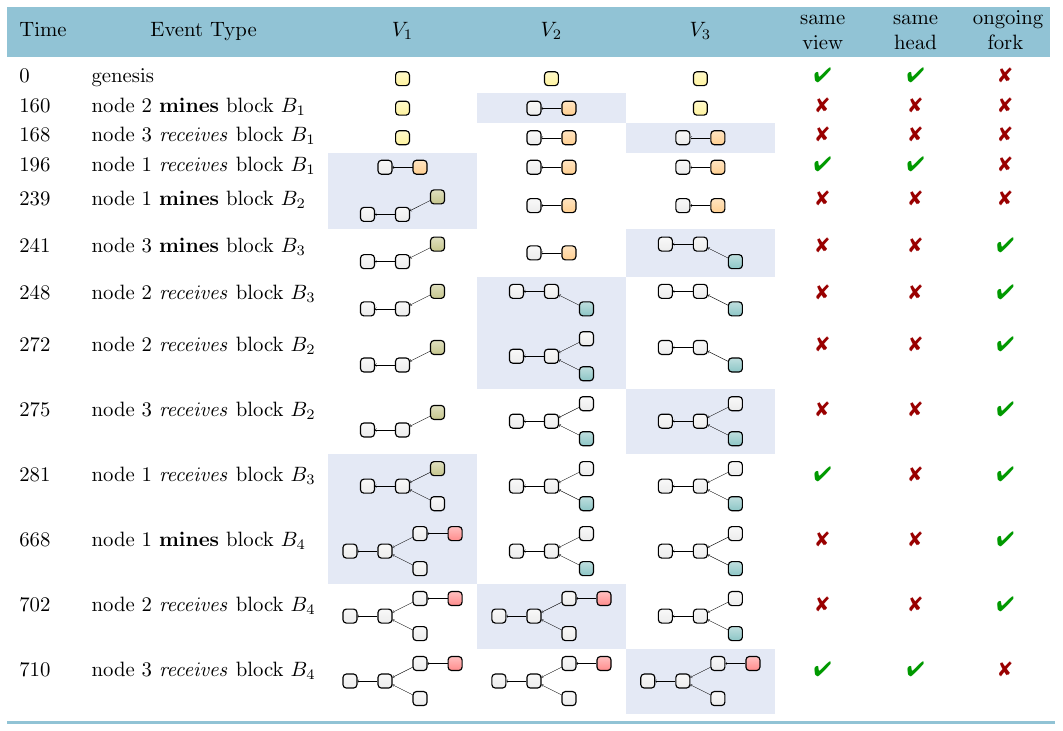}}
\caption{An example of the evolution of blockchain network of \Cref{fig:networkandblockchainfigure}. Each row corresponds to an event: it displays its time $t$ of occurrence and its type (mining or propagation). Furthermore, we display the view $V_i$ of each node $i \in \{1,2,3\}$ after the event, and note whether the nodes' views and/or heads are consistent across the network and whether a fork is ongoing. Events related to the dissemination of transactions are not included for brevity. 
}\vspace{-0.3cm}
\label{tab:evolution}
\end{table*}

\subsection{Blockchains as Peer-to-Peer Networks} \label{sec:network}
Each node $n \in \N$ has incoming and outgoing connections to other protocol-running nodes. The resulting \spemph{peer-to-peer network} can be represented as a graph where the vertices represent nodes and edges represent connections -- see \Cref{fig:networkfigure} for an illustration. Not every node is necessarily connected to all other nodes, however it is typically assumed that from each node a path to every other node exists in the graph. If this is not the case, then there is an ongoing network \spemph{partition}.\par

At each time $t\ge0$, each node $n\in \N$ is aware of a set $V_n^t \subseteq \N$ of blocks: we call $V_n^t$ the \spemph{view} of node $n$ at time $t$. The genesis block $g$ is the only block that all nodes are aware of at time $0$, i.e., $V^0_n=\{g\}$ for all $n\in \N$. 
In addition to $V_n^t$, each node $n\in \N$ is also aware of transactions that have not been included in any block at time point $t>0$. This information is stored in the \spemph{memory pool}, denoted by $M_n^t$, for $n\in \N$ and $t>0$. Due to the distributed nature of the network and the presence of network latency, there must exist points in time for which different nodes are aware of different sets of blocks or different information, i.e., there exist $t>0$ and nodes $n,m\in \N$ such that $V^t_n\neq V^t_m$ or $M^t_n\neq M^t_m$.\par
At any time $t$, each node $n\in\N$ that can create blocks has to decide which block in $V^t_n$ to extend. This block is called the \spemph{head}, and is represented by the variable $H^t_n$. A function $f$ that selects a head from a view is called a \spemph{fork-choice rule}.
A \spemph{network fork} is any period during which at least two (protocol-following) nodes have \spemph{incompatible} blocks as heads. Here, we mean by \qt{incompatible blocks} two blocks $B$ and $B'$ such that neither is in the chain of the other, i.e., $B \notin C(B')$ and $B' \notin C\(B\)$. The term \qt{fork} is also commonly used in practice to refer to protocol changes. That is, if the protocol is changed from $\Pi$ to $\Pi'$ and blocks created under $\Pi'$ are still considered valid by $\Pi$, then this is referred to as a \emph{soft fork}. If not, the change is called a \emph{hard fork}. \par
Due to network forks, a block $B$ can be \spemph{orphaned}, which occurs if at some point of time $t$, there is no $n \in \N$ such that $B \in C(H^t_{n})$. 
We say that a block $B$ is \spemph{overturned} by node $n$ at time $t$ if $B \in \lim_{\epsilon \downarrow 0} C(H^{t-\epsilon}_n)$ and \mbox{$B \notin C(H^t_n)$}. 
For example, in \Cref{tab:evolution}, a network fork occurs from time 241 until time 710. Block $B_3$ (the teal block) is overturned by node 2 at time 702 and by node 3 at time 710. In practice, blockchain users need either a formal or heuristic notion of \spemph{finality} -- i.e., a notion of when a block can be assumed to not be overturned. For example, an online retailer will need to decide when a block that contains a payment is safe enough from being overturned to  dispatch the order.

\setcounter{example}{0}
\begin{example}[Bitcoin] In Bitcoin, the fork-choice rule prescribes to select the block $B$ with the highest accumulated proof-of-work, i.e., 
\[f\(V\) = \argmax_{B \in V} \sum_{B' \in C(B)} W\(B'\).\]
In case of ties, the block seen first is preferred. This can lead to soft forks that persist even when all nodes have the same view, as illustrated in Table~\ref{tab:evolution}. In \cite{Ey14}, it was suggested that adversarial behavior can be discouraged by using \emph{uniform tie breaking} -- whenever a node learns of a new block that has as much proof-of-work as its head, it adopts the new block as its head with probability $0.5$. As further discussed in \cite{Sa17}, this can have either a positive or negative effect on attackers, depending on how well-connected they are within the network.

For finality, Bitcoin users typically use the six-confirmations rule \cite{sirer2016online}: i.e., a block $B$ is considered final by $n$ at time $t$ if there exists a $B' \in V^t_{n}$ such that $B = P^6(B')$.
\end{example}
\begin{example}[Quorum] In the Istanbul-BFT protocol used by Quorum, blocks are added to the blockchain if they are confirmed by more than $2/3$ of the voters. In particular, let, for any block $B \in \N$ and private key $i \in \{1,\ldots,k\}$, ${\bf 1}_{M_n^t}(i,B)$ equal $1$ if $M_n^t$ contains a \qt{commit} message for $B$ signed with $i$, and 0 otherwise. Then block $B$ is considered a valid block by node $n$ at time $t$ if
\[\textstyle \sum_{i = 1}^k {\bf 1}_{M_n^t}(i,B) \geq \(2/3\) k.\]
This is also the finality rule: i.e., in Quorum, all blocks are either both valid and finalized, or neither. This is true for many other BFT protocols as well.
\end{example}

\subsection{Actions, Strategies, and Utilities} \label{sec:incentives}

Based on the above, the \spemph{state} $\rstate_t$ of the blockchain at timepoint $t\ge0$ is a vector
\begin{equation}\label{eq:state}\rstate_t:=\(V_n^t,H^t_n,M_n^t,r_n^t\)_{n\in \N}.\end{equation}
When time is not relevant, we simply write $\rstate$ instead of $\rstate_t$. The \spemph{state space}, i.e., the set of all possible states $X_t$ will be denoted by $\X$. \par
Transitions to new states typically occur through operations or \spemph{actions} performed by the nodes. Each protocol $\Pi$ has its own set of actions, and conditions under which they are available.
We denote by $\A$ the set of all possible actions allowed by the protocol. 
Let a \spemph{strategy} $s: \states \to \A$ be a function from the state space to the set of actions -- i.e., during execution, a node uses its strategy to select which action to take given the state of the system (where `waiting' can be an action). Let $\S$ denote the set of all possible strategies. Particularly relevant to our presentation is the \spemph{default strategy} or the strategy that prescribes to \emph{follow-the-protocol} as referenced, which we will denote by $d$. We will call any other strategy $s\in \S$ with $s\neq d$ a deviant or \spemph{adversarial} strategy. Also, we will refer to nodes that follow $d$ as \spemph{honest} and to nodes that do not follow $d$ as {adversarial} nodes.
If a node takes an action, then the system state is changed after a random delay. Let $A_t$ be the set of completed actions until time $t$. This imposes a probability measure on system executions, i.e., given a strategy profile $S = \times_{n \in \N} \, s_n$, then for all $y,z \in \states$ and $t,\delta \geq 0$ we should be able to determine $\P_S(\rstate_{t+\delta} = y \,|\, \rstate_{t} = z)$.
Each agent $n\in\N$ has a \spemph{utility} function $u_n: \states \to \R$, which indicates their satisfaction with a given network state, and a utility function $u'_n : \A \to \R$, which indicates their satisfaction with a given action. Given an initial state $x_0$, the \spemph{long-term average utility} $v_n(S,x_0)$ with a given strategy profile $S$ is then given by
$$
v_n(S,x_0) = {\mathbb E_S} \left(\lim_{t \to \infty} \frac{1}{t} \left(  \int_{0}^t u_n(\state_{\tau}) d\tau + \sum_{a \in A_t} u'_n(a) \right) \right).
$$
\par
In the present study, rather than analyzing individual strategic behavior, we mainly focus on collective protocol performance and blockchain properties. To formalize these notions, we define a \spemph{property} or \emph{feature} of a blockchain protocol $\Pi$ as a function $\F_{\Pi}:\states\to\{0,1\}$, with the following values
\[\F_\Pi\(X_t\)=\begin{cases}1, & \text{if $\Pi$ satisfies property $\F_\Pi$ at state $X_t$},\\0, & \text{otherwise.}\end{cases}\]
Additionally, we define aggregate utilities or \spemph{performance measures} of protocol $\Pi$ as functions $U_{\Pi}:\states\to\mathbb R$. We say that a performance measure is \emph{positive} when higher values indicate better performance, e.g., throughput rate and collective profits, and \emph{negative} when lower values indicate better performance, e.g., communication complexity and operational costs.\par

\setcounter{example}{0}
\begin{example}[Bitcoin] The core actions that any node $n$ in Bitcoin can perform are \emph{block creation}, \emph{block propagation}, \emph{transaction creation}, and \emph{transaction propagation}. Of these, {block creation} takes an amount of time that is approximately exponentially distributed \cite{rosenfeld2011analysis,decker2013information} with a mean that is determined by the node's processing power. The time needed to create transactions is negligible. Block and transaction propagation times depend on the network latency, the node's bandwidth, connectivity, and the message size. More generally, \emph{block validation} can be included as a separate action that consumes processing power, to model the profitability of so-called SPV mining (i.e., mining a block without validating its transactions). Also, if a node $n$ represents a mining pool, then the \emph{entering} or \emph{exiting} of $n$ by other nodes can be considered as actions. 
\end{example}
\begin{example}[Quorum] Instanbul-BFT has the same core actions as Bitcoin, but also includes the propagation of protocol messages, e.g., prepares and commits. Unlike Bitcoin, block creation is nearly instantaneous, but a node's block is only valid if it has been selected as the block proposer (e.g., via a round-robin scheme) for the current \emph{round}. Under certain conditions, e.g., if a block proposer waits too long before proposing a block, the other nodes can request a \emph{round change}. If a supermajority of nodes agree on a round change, or sign off on a block, the next round begins.
\end{example}

Throughout the text, we will frequently use the following terms (and abbreviations):
\begin{itemize}[leftmargin=*]
\item \textbf{Proof of Work} (PoW): As already mentioned, this refers to Nakamoto consensus \cite{Na08} in which nodes, also called miners, gain the right to participate in the block creation process by providing solutions to a computationally difficult and energy-consuming cryptographic puzzle. 
\item \textbf{Proof of Stake} (PoS): Often called Virtual Mining \cite{Be16}, PoS emulates the above process but saves on energy waste by requiring from participating nodes to provide proof of \qt{virtual} resources such as the platform's native tokens.
\end{itemize}
We will call a protocol \spemph{permissionless} if the consensus-critical resources are not inherent to the blockchain (e.g., processing power in PoW). We will a protocol \spemph{semi-permissionless} if the consensus-critical resources are inherent to the blockchain, but freely divisible and transferable (e.g., tokens in PoS). We will call a protocol \spemph{permissioned} if the consensus-critical resources are inherent to the blockchain and indivisible (e.g., the private keys in Quorum).

We are now ready to define the 5 axes of the PREStO framework and describe their subcategories.

\section{Optimality}
\label{sec:optimality}
\emph{Optimality} is the most basic property of a protocol, and generally refers to whether the protocol is optimal within its operational scope. In our setting, it concerns the question:
\begin{itemize}
\item[Q:] \emph{Under normal conditions, does the protocol provide its core functionality in an optimal way?}
\end{itemize}
By \qt{normal conditions}, we mean that nodes do not act strategically or maliciously, and that there are no capacity constraints. However, we do consider network latency and nodes going offline. \qt{Core functionality} primarily refers to the functionality of any distributed database, i.e., to correctly read and write to the database. 
However, some protocols also provide additional functionalities, e.g., a broader notion of transaction types, or a higher level of privacy. 

\subsection{Liveness and Safety}
Since blockchains are essentially data structures, they must adequately perform the \emph{read} and \emph{write} operations that are required of any database. We focus on the data in the finalized blocks of the chain, since non-finalized blocks can be overturned, \cite{sirer2016online}. The \qt{write} operation then consists of adding a transaction to a finalized block on the chain. The \qt{read} operation consists of observing that a transaction has made it into a finalized block on the chain. \par
The ability to write and read correctly is formalized through the notions of \emph{liveness} and \emph{safety}. A {liveness} fault means that a node is unable to write to the blockchain. A {safety} fault means either that two honest nodes see different results when reading the database, or that a single node sees different results when reading the database at different times.\footnote{These two types of safety faults are practically equivalent: if two nodes see different results, then either the network remains permanently forked, or at least one of them will read a different value at some point in the future.} This informs our two general definitions of liveness and safety below.
\begin{definition}[Liveness]\label{def:live}
We say that a protocol $\Pi$ is \emph{live} if from any state $\rstate \in \states$, any protocol-following node $n \in \N$ can take a sequence of actions $a_1,a_2,\ldots,a_m$ that will lead to a valid transaction being added to a final block in its view.
\end{definition}
\begin{definition}[Safety]\label{def:safety}
We say that a protocol $\Pi$ is \emph{safe} if the following holds: any protocol-following node $n \in \N$  who considers a block $B$ as final at some time point $T \geq 0$, will also consider $B$ as final at any time point $t > T$.
\end{definition}
In practice, most protocols satisfy these properties only under certain conditions. In particular, most require that the honest nodes control a given fraction of the consensus-critical resources. For example, Bitcoin is safe only if the honest nodes are over $50\%$-strong in terms of processing power. Even then, the property of safety is guaranteed only in a probabilistic sense.\footnote{However, this probability can be made arbitrarily high by increasing the number of confirmations required to make a block final.} Quorum is live only if the honest and not permanently offline nodes are at least $\frac{2}{3}$-strong, and safe only if the adversarial nodes are less than $\frac{2}{3}$-strong in terms of authority.

During a network partition, protocols can either satisfy liveness or safety, but not both -- this is known as the CAP theorem \cite{gilbert2002brewer}.
Different protocols resolve this trade-off in different ways. Liveness-oriented protocols such as Nakamoto consensus allow the chain to fork by providing an unambiguous rule of how to resolve such forks when the partition ends, e.g., the longest-chain rule. Safety-oriented protocols such as Tendermint \cite{Kw14} and most other Byzantine fault tolerant protocols  \cite{Cac16,Vu15} require that a (super)majority of participants sign off on each block. This means that during a network partition, at least one of the branches of the chain stops growing. In other settings, different branches of the chain can grow during a fork, but only one of these branches can \emph{finalize} blocks. Examples include a traditional proof-of-work chain with Casper the Friendly Finality Gadget as an overlay (`hybrid' Casper) \cite{Bu19,buterin2017casper}. It is also possible for protocols to guarantee neither liveness nor safety, e.g., Tangaroa \cite{Ca17}.

In the scientific literature, the definitions and terminology used for safety and liveness properties vary. Algorand \cite{Gi17} provides its own definitions of liveness (new transactions can be added to the final part of the blockchain) and safety (if a node accepts a transaction as final, then it will continue to do so). In \cite{Ga15,kiayias2017ouroboros,david2018ouroboros,Pa17a}, safety is called \emph{persistence} -- in turn, persistence and liveness can be shown to follow from  three properties: \emph{common prefix}, \emph{chain quality}, and \emph{chain growth}. For the Snow White \cite{bentov2016snow,daian2016snow} and Tortoise and Hares \cite{bentov2017tortoise} protocols, chain growth, chain quality, and \emph{consistency} are considered. Here, consistency is a combination of common prefix and \emph{future self-consistency}. In \cite{Ba17}, the properties of \emph{validity} and \emph{agreement} are discussed as liveness properties, whereas \emph{integrity} and \emph{total order} are used for safety.

Typically, liveness and safety are proven for a specific protocol through a bespoke mathematical proof. However, under some assumptions on the actions in the protocol, general proof techniques may be available, e.g., model checking \cite{biere2003bounded}. 

\subsection{Transaction Scope}

Some protocols offer fundamentally different types of transactions than others. For example, Bitcoin only supports monetary transactions, which allows for the entire \qt{state} of the system to be described using unspent transaction outputs (UTXOs). However, protocols that support smart contracts (e.g., Ethereum \cite{Bu14online}) require that the clients also store the internal variables of the contracts \cite{Sou18}. This may have an impact of efficiency (see also \Cref{sec:efficiency}), both via reduced throughput due to slower transaction processing, and potentially less straightforward scalability (``state sharding'' \cite{zilliqawp}).

\subsection{Privacy}
The choice to put data on a blockchain instead of a centralized database has implications for privacy. On one hand, permissionless blockchains such as Bitcoin do not require identity management, thus favoring privacy. On the other hand, the entire history of transactions is publicly accessible, which may allow for de-anonymization. In fact, Bitcoin transactions may be better described as pseudonymous than as anonymous \cite{Cat16}. Cryptographic techniques that improve privacy, e.g., zero-knowledge proofs \cite{goldwasser1989knowledge} or ring signatures \cite{rivest2001leak}, are available, although they may impose additional computational overhead and therefore impact efficiency. Furthermore, usage pattern analysis can lead to user de-anonymization even in privacy-minded platforms such as Zcash \cite{kappos2018empirical}.

\section{Stability}
\label{sec:stability}
Since intended behavior cannot be enforced in decentralized settings, one of the core tasks of consensus protocols is to properly incentivize agents to behave appropriately. This will enable the network to reach an outcome that is both stable and desirable. 
Importantly, stability does not imply optimality; instead, it is concerned with the question:
\begin{itemize}
\item[Q:] \emph{Does the protocol incentivize the intended behavior? Is implementing and following-the-protocol the best possible strategy for participating and prospective nodes?} 
\end{itemize}
Game theory and traditional economics provide numerous tools to analyze this setting. Yet, 
as consensus protocols become more elaborate, the incentives and the required stabilizing mechanisms also become more complicated. These issues are discussed separately below.

\subsection{Incentive Compatibility}
At its core, incentive compatibility entails that it is in the participants' best interest to \emph{follow-the-protocol}. In concrete terms, and using the notation of \Cref{sec:model}, this means that the default strategy profile $\times_{n\in  \NN}d$  is a Nash equilibrium, \cite{Ne44,Na50}. An equilibrium is an outcome that is optimal from the perspective of all decision makers involved. Formally,
\begin{definition}[\emph{Incentive Compatibility}] \label{def:Nash} Let $\Pi$ denote a protocol with active nodes $\NN \subset \N$, set of available strategies $\S$, and long-term average utility functions $v_n$ as defined in \Cref{sec:model} for each $n\in  \NN$. Also, let $d \in \S$ denote the \emph{follow-the-protocol} strategy, $D = \times_{n \in \NN} \; d$ the strategy profile where all nodes follow $d$, and $D_{n,s}$ the profile where all nodes follow $d$ except node $n$ who follows $s \in \S$. Then, $\Pi$ is incentive-compatible, if 
\[v_n\(D, x_0\)\ge v_n\(D_{n,s}, x_0\), \quad \text{for all } s\in \S,n \in \NN, \] and initial states $x_0 \in \states$. In words, $\Pi$ is incentive-compatible, if \emph{given that} all other nodes follow the protocol, then it is optimal for an entering (or existing) node to also follow the protocol.
\end{definition}

This definition relies on some assumptions that are not always satisfied in practice. It assumes that first, each node can take as given that all other agents do follow the protocol (strategy $d$) and second, that all agents are \emph{rational}, i.e., utility maximizers. Also, it requires that utility functions are known for each node. Although this set of assumptions may seem restrictive, it is an essential first step in protocol design to establish stability of a protocol within this {vanilla} setting. It is within the scope primarily of robustness and to a lesser extent of persistence to explore what will happen if these assumptions are violated, cf. \Cref{sec:robustness,sec:persistence}. 

\setcounter{example}{0}
\begin{example}[Bitcoin]
Nash equilibria in Bitcoin mining are discussed in \cite{Ey14} and \cite{Ki16}. These works show that the Bitcoin protocol is an equilibrium only if a potential adversary (which could just be a rational node) is not too strong in terms of precessing power. \cite{Kr13} find that there is a multitude of symmetric equilibria in the Bitcoin protocol. Still, the default strategy prevails by a \emph{focal-point} argument \cite{Sc05,My07}. \cite{Ey14} and \cite{Sa17} show that if nodes are at least $\alpha$-strong, where $\alpha$ depends on their connectivity, then they are incentivized to follow the adversarial \emph{selfish mining} strategy. \cite{Nay16} combine selfish mining -- a consensus-layer attack -- with an eclipse attack -- a network-layer attack -- to augment the rewards of selfish mining. 
Despite their theoretical plausibility, such attacks have not been recorded in practice. 
\end{example}
Based on the above, the task of the blockchain architect is to design the consensus protocol in a way to induce the desired behavior in practice. Differences between intended and observed behavior should be addressed at this point. The theoretical discipline that models and studies such settings is that of \emph{mechanism design} \cite{Hu06, My07, Ba11}. Applied in the blockchain context, it aims to determine the rules of the protocol in such a way that individual incentives are perfectly aligned with societal goals. 

The notion of incentive compatibility can be seen beyond just Nash equilibria. For example, depending on whether the majority is controlled by a single entity or not, one may discern between \emph{strong} and \emph{weak} incentive compatibility \cite{Bo16}.
In practice, a consensus protocol of a public, permissionless blockchain needs to properly incentivize rational agents to perform the following actions:
\begin{itemize}[leftmargin=*]
\item Participation: acquire protocol resources, e.g., bandwidth.
\item Operations: perform core and auxiliary tasks such as proposal and creation of blocks, message propagation, transaction validation and execution, data storage, etc. \cite{Cat16}.
\item Applications: use the native cryptocurrency or blockchain related applications (``Dapps'').
\end{itemize}
Although they are integral to viability of existing blockchains, not all of these actions are properly incentivized, and miners' incentives may be at odds with the underlying protocol \cite{So18}. Additional concerns stem from the tension between short-term and long-term incentives \cite{La15}. In \cite{Re14}, a consensus protocol is proposed that motivates both ownership and participation, and which aims to develop blockchains for social interaction. In \cite{Ga18}, it is shown that the core economic motives for miners -- transaction fees and block rewards -- are also inherent to the security of PoS protocols. Finally, recent works suggest reputation systems as possible solutions to improve the incentive mechanisms of consensus protocols \cite{No17,Le19}.\par
The diversity of entities that are involved in the blockchain ecosystem introduces additional complexity. Different groups ranging from investors, developers, token holders to participating nodes and end users often have conflicting incentives. This implies that apart from the need to incentivize certain operations, like the ones mentioned above, the blockchain protocol also needs to align potentially conflicting incentives of these groups. Similar concerns emerge in blockchain-based markets or applications that entail the interaction between infrastructure operators, cyber-security providers, entrepreneurs, and end users in a trustless environment \cite{Fe18}.\par
The theory on social choice and public goods provides insight into misaligned blockchain incentives \cite{Sc99}. A notable instance is captured by the \emph{free-rider} or \emph{pass-the-bucket} problem \cite{Ba52, St16}. In simplified terms, it states that rational agents who benefit from the existence of a public good -- in this case, the blockchain -- may shift responsibility for its creation to their peers. In the resulting equilibrium, the public good is not created, to everyone's detriment. In public, permissionless blockchains, this translates to nodes moving costly tasks to other nodes, leading to an improper functionality of the blockchain ecosystem and a deviation from its intended outcome. 

\setcounter{example}{0}
\begin{example}[Bitcoin]
To explain the lack of observed selfish mining incidents, \cite{Bad18} suggest natural incentives (i.e., mining rewards) and the high monetary value of Bitcoin as mitigating factors \cite{Ga13}. Yet, \cite{Ey15} identify incentives for attacks between miners and argue that the prevailing practice of not engaging in these attacks is fragile and if broken will lead to equilibria with dire consequences for the blockchain ecosystem.\par
In \cite{Ba12}, it is argued that the Bitcoin reference protocol disincentivizes the propagation of information. In \cite{At16}, it is demonstrated that the active usage of Bitcoin remains low, and largely restricted to speculation and illegal activity --- this has a negative impact on the stability of Bitcoin's exchange rate with fiat currencies, an issue also discussed in \cite{Kr13}. In \cite{Zh17}, a scheme to incentivize miners to promptly propagate any blocks that they know of is proposed as a way to effectively defend against certain adversarial strategies.\par
Many blockchain platforms include transaction fees that are paid by the transaction creator to the node that creates a block that includes their transaction. Transaction fees also impact the incentives of Bitcoin users. Currently, Bitcoin transaction fees are low, yet non-zero \cite{Kr13}. However, in Turing-complete environments (e.g.,  Ethereum), transactions typically require more computation, which makes these environments particularly vulnerable to network-layer attacks \cite{Lu15a}. Based on the resource utilized most, there are three main sources of cost for the miners: network, computation and storage. \cite{Ch18} propose a fee-paying scheme for memory consumption that is typical to cloud storage services. In contrast to the currently deterministic transaction fees, \cite{Gj16} put forward the idea of random payments to incentivize risk-neutral miners. \cite{Ca16} analyze the future of Bitcoin mining, when mining rewards will have diminished, and conclude that if transaction fees are the only incentive, then selfish mining will be profitable for miners with arbitrarily small wealth. In the case that transaction fees can be further reduced, \cite{Chi17} argue that cryptocurrencies have the potential to become viable alternatives to retail payment schemes.
\end{example}

\paragraph*{Protocol Resources}
Protocol stability is tightly linked to the way that participating nodes acquire and increment their resources, which is starkly different between, e.g., PoW and PoS. In PoW protocols such as Bitcoin, computational (CPU) power is the consensus-critical resource. This implies that the costs for participating nodes are mainly electricity and investment in mining equipment \cite{Dh18}. These resources can be acquired in fiat currencies, yet the mining rewards are distributed in the native cryptocurrency (Bitcoins).\par
PoS protocols generate different dynamics. Virtual miners acquire their resources by converting fiat currency to the native cryptocurrency, which they then use as a proof to participate in the consensus mechanism. Mining rewards are again distributed in the native currency, however, in this case, the rewards naturally contribute to the protocol resources. This creates conflicting motives for staking nodes, since wealth and resources coincide and may lead to unpredictable inflation or disincentives to use or spend one's stake. Along with latent economic effects that affect cryptocurrency prices \cite{Kok20}, these observations call for a re-evaluation of the economics of different protocols through the lens of novel macroeconomic tools. Integral are the questions about the distribution of resources, the corresponding entry barriers, and market dynamics -- perfect or oligopolistic competition -- that they induce \cite{Ar19,Leo19}.

\subsection{Decentralization}
\label{sub:decentralization}
Decentralization lies at the core of blockchain design philosophy and is therefore integral for its long-term survival and sustainability \cite{Lu17,Leo19}. However, existing data shows that centralization plagues PoW (and PoS) cryptocurrencies of both high and low market values \cite{Fi17, Brj18}. Miners join centralized pools which efficiently distribute mining rewards among their participants. This is known to reduce the extreme variance of mining returns that discourages solo miners, \cite{rosenfeld2011analysis,Sc17,szalachowski2019strongchain}. Yet, the operation of mining pools introduces unpredictable dynamics in the consensus mechanism and incentivizes miners (or protocol participants) to behave dishonestly, especially under high transaction loads, and destabilize the system \cite{Le15, No17}. For instance, staking pools -- the equivalent of mining pools in PoS protocols -- can potentially evolve to become institutions with arbitrary power over their cryptocurrency \cite{Fa18,Brj18}.
\par
In conventional market economics, market concentration is measured by the \emph{Herfindahl-Hirschman Index} (HHI), \cite{Rh93}.  In general, the HHI is defined as the sum of the squares of the market shares of the firms within the industry, where the market shares are expressed as percentages. As such, it can range from 0 to $10,000$, with higher values indicating larger concentration\footnote{The USA Department of Justice considers a market with an HHI of less than 1,500 to be a competitive marketplace, an HHI of 1,500 to 2,500 to be a moderately concentrated marketplace, and an HHI of 2,500 or greater to be a highly concentrated marketplace. However, these thresholds refer to oligopolistic markets and should be much lower when studying decentralization in blockchains.}. In the blockchain context, it can be used to study the concentration of protocol resources. For a state $\state$ of protocol $\Pi$ with active nodes $\NN \subset \N$, and distribution of consensus-critical resource fractions $\(p_{n}\)_{n\in \NN}$, the HHI is given by 
\[\textstyle\operatorname{HHI}_{\Pi}\(\state\):=\sum_{n\in \NN}\(p_{n}\cdot100\%\)^2\]
The HHI is used in the context of antitrust management and also in applied social and political sciences to measure the concentration of political power \cite{US19online, Le11}.

\begin{example}[Bitcoin \& Ethereum]
\Cref{tab:decentralization} shows the estimated distribution of mining power between the top 10 Ethereum mining pools (or accounts) by number of blocks. 

\begin{table}[!htb]
\centering
\vspace{0.2cm}
\setlength{\tabcolsep}{5.8pt}
\renewcommand{\arraystretch}{1.2}
\begin{tabular}{LlrclR}
\rowdark
& \multicolumn{2}{@{}c@{}}{\sc \b Bitcoin}&& \multicolumn{2}{@{}c}{\sc \b Ethereum}\\
\rowdark
& Entity (Pool) & Blocks \% && Entity (Pool) & Blocks \% \\\\[-0.3cm]
\rowlight
1. & BTC.com & $20.1\%$ && Ethermine & $26.5\%$ \\
2. & AntPool & $14.5\%$ && Sparkpool & $24.5\%$\\
\rowlight
3. & F2Pool & $13.1\%$ && F2Pool\_2 & $11.8\%$\\
4. & Slushpool & $8.8\%$ && Nanopool & $11.2\%$\\
\rowlight
5. & Poolin & $8.8\%$ && MiningPoolHub\_1 & $5.4\%$\\
6. & ViaBTC & $8.3\%$ && Address\_1 & $2.3\%$\\
\rowlight
7. & BTC.TOP & $6.1\%$ && Address\_2 & $1.7\%$\\
8. & BitFury & $4.9\%$ && DwarfPool 1& $1.7\%$\\
\rowlight
9. & BitClub Network & $1.7\%$ && zhizhu.top & $1.3\%$\\
10.& Bitcoin.com & $1.4\%$ && firepool & $1.2\%$\\
\rowdark
\multicolumn{2}{r}{\b Total:} & \b $87.7\%$ && & $87.6\%$ \\
\rowdark
\multicolumn{2}{r}{\b HH Index:} & \b1075.7 && & \b1610.5 \\\\[-0.3cm]
\bottomline
\end{tabular}\\[0.2cm]
\caption{Concentration of mining power for the Bitcoin and Ethereum blockchains (as of 07 June 2019) and calculation of the Herfindahl Hirschman Index (HHI). Sources: \href{https://www.blockchain.com/en/pools}{blockchain.com} and \href{https://etherscan.io/stat/miner?blocktype=blocks}{etherscan.io}.}\vspace{-0.2cm}
\label{tab:decentralization}
\end{table}
The figures indicate a more decentralized market for Bitcoin than for Ethereum. Similar calculations indicate even higher centralization for smaller PoW platforms, for which $51\%$ attacks -- distributions in which a single entity owns more than $50\%$ of the resources -- are a reality \cite{Al18online,Je19online,Ge16}. These figures echo the concerns that the current structure of some of the major blockchain platforms is prone to centralization \cite{Ge14, Ap16, Arn18, Leo19}. \par
From a stability perspective, \cite{Nay16} argue in favor of dispersing mining power since the \emph{follow-the-protocol} strategy ceases to be a Nash equilibrium if a single node becomes too strong. Incentives to derive short-term profits from attacks on mining pools threaten the long-term viability of Bitcoin and negatively impact the Bitcoin ecosystem \cite{La15}. \cite{Jo14} show that pool size and computational power are the main criteria when deciding whether to launch a network-level attack against a mining pool. These concerns are not only relevant to Bitcoin, but to other PoW blockchains as well \cite{Ge16}.  
\end{example}

Ideally, nodes should have no motive to band together at all. This informs the following definition:

\begin{definition}[\emph{Perfect Decentralization}] Let protocol $\Pi$ in any state $X \in \states$ consist of nodes $\NN \subset \N$, such that each node \mbox{$n \in \NN$} controls a fraction $p_{n}$ of the consensus-critical resource. Let the state $X_{nm}$, for $n,m \in \NN$, be the same state as $X$ except with nodes $n$ and $m$ merged, and their resources combined. Also, let $d \in \S$ denote the \emph{follow-the-protocol} strategy and $D = \times_{n \in \NN} \; d$ the strategy profile where all nodes follow $d$. Then a protocol satisfies \emph{perfect decentralization} if
$$
v_{n}(D, X) \geq v_{n}(D, X_{nm}) \text{ for all } n,m \in \NN.
$$
\end{definition}
Such a definition depends strongly on the utility functions: e.g., in Bitcoin, banding together always reduces the reward variance, but when the pools get too strong, trust in the system is undermined and Bitcoins will lose value against other (crypto)currencies. For example, the mining pool GHash.IO was forced to take action to reduce their pool size after they surpassed the $50\%$ mark \cite{ghash}.

\begin{figure}[!htb]
\vspace{-0.2cm}\centering
\begin{tikzpicture}
\clip (-0.2,0) rectangle (8.0,8.0);

\Vertex[x=0.4,y=7.5,size=0.2,style={color=blue!20}]{m1}
\Vertex[x=0.2,y=6,size=0.4,style={color=blue!20}]{m2}
\Vertex[x=1.3,y=6.1,size=0.1,style={color=blue!20}]{m3}
\Vertex[x=2,y=7.4,size=0.5,style={color=blue!20}]{m4}
\Vertex[x=3,y=6.8,size=0.3,style={color=blue!20}]{m5}
\Vertex[x=2.4,y=5.2,size=1,style={color=blue!30}, label=Pool 1]{pool1}
\Edge[,lw=0.5,bend=8.531,Direct](m1)(pool1)
\Edge[,lw=0.5,bend=-8.531,Direct](m2)(pool1)
\Edge[,lw=0.5,bend=-8.531,Direct](m3)(pool1)
\Edge[,lw=0.5,bend=8.531,Direct](m4)(pool1)
\Edge[,lw=0.5,bend=8.531,Direct](m5)(pool1)

\Vertex[x=5.1,y=7.5,size=0.2,style={color=blue!20},position=below]{m6}
\Vertex[x=4.6,y=6.3,size=0.4,style={color=blue!20}]{m7}
\Vertex[x=6.3,y=6.5,size=0.1,style={color=blue!20}]{m8}
\Vertex[x=7,y=7.4,size=0.5,style={color=blue!20}]{m9}
\Vertex[x=7.6,y=6,size=0.5,style={color=blue!20}]{m10}
\Vertex[x=6.3,y=5.2,size=1.2,style={color=blue!30}, label=Pool 2]{pool2}
\Edge[,lw=0.5,bend=8.531,Direct](m6)(pool2)
\Edge[,lw=0.5,bend=-8.531,Direct](m7)(pool2)
\Edge[,lw=0.5,bend=-8.531,Direct](m8)(pool2)
\Edge[,lw=0.5,bend=8.531,Direct](m9)(pool2)
\Edge[,lw=0.5,bend=8.531,Direct](m10)(pool2)
\Text[x=4.8,y=5.2, width=3cm,fontsize=\scriptsize]{Mining/Staking Pools}
\Text[x=1.85,y=6.8, width=3cm,fontsize=\scriptsize]{Miners}
\Text[x=6.25,y=6.9, width=3cm,fontsize=\scriptsize]{Miners}
\Text[x=2.85,y=6.8, width=3cm,fontsize=\scriptsize]{Stakers}
\Text[x=7.15,y=6.9, width=3cm,fontsize=\scriptsize]{Stakers}

\Plane[x=3,y=-1, width=1,height=11,color=white, style={dashed,color=white}, NoBorder, opacity=0]
\Vertex[x=2,y=3.5,shape=rectangle, label=B1, opacity=0.7]{b1}
\Vertex[x=3,y=3.5,shape=rectangle, label=B2, opacity=0.7]{b2}
\Vertex[x=4,y=3.5,shape=rectangle, label=B3, opacity=0.7]{b3}
\Vertex[x=5,y=3.5,shape=rectangle, label=B4, opacity=0.7]{b4}
\Vertex[x=6,y=3.5,shape=rectangle, label=B5, opacity=0.7]{b5}
\Vertex[x=7,y=3.5,style={color=white},opacity=0,label=$\dots$]{b6}
\Vertex[x=1,y=3.5,style={color=white},opacity=0,label=$\dots$]{b0}
\Edge[,lw=0.5,Direct](b1)(b0)
\Edge[,lw=0.5,Direct](b2)(b1)
\Edge[,lw=0.5,Direct](b3)(b2)
\Edge[,lw=0.5,Direct](b4)(b3)
\Edge[,lw=0.5,Direct](b5)(b4)
\Edge[,lw=0.5,Direct](b6)(b5)

\Edge[,lw=2,bend=10,Direct](pool1)(b3)
\Edge[,lw=2,bend=-10,Direct](pool2)(b4)
\Text[x=7.7,y=3.7,fontsize=\scriptsize]{Block}
\Text[x=7.7,y=3.4,fontsize=\scriptsize]{chain}

\Vertex[x=0.9,y=1.7,size=0.4,style={color=red!20}]{u1}
\Vertex[x=1.4,y=0.3,size=0.1,style={color=red!20}]{u2}
\Vertex[x=0.6,y=1,size=0.5,style={color=red!20}]{u3}
\Vertex[x=3.5,y=0.2,size=0.2,style={color=red!20}]{u4}
\Vertex[x=5,y=0.3,size=0.4,style={color=red!20}]{u5}
\Vertex[x=6,y=0.8,size=0.3,style={color=red!20}]{u6}
\Vertex[x=3,y=1.4,size=1.5,style={color=red!30}, label={Fin-tech}]{fin1}
\Edge[,lw=0.5,bend=8.531,Direct](u1)(fin1)
\Edge[,lw=0.5,bend=-8.531,Direct](u2)(fin1)
\Edge[,lw=0.5,bend=8.531,Direct](u3)(fin1)
\Edge[,lw=0.5,bend=8.531,Direct](u4)(fin1)
\Edge[,lw=0.5,bend=-8.531,Direct, label=Users](u5)(fin1)
\Edge[,lw=0.5,bend=-8.531,Direct](u6)(fin1)
\Edge[,lw=2.0,bend=-10,Direct](fin1)(b3)

\Vertex[x=7,y=2,size=1.3,style={color=gray!20},label=Services]{mon1}
\Vertex[x=0.4,y=4.6,size=1.1,style={color=gray!20}, label=Services]{mon2}
\Edge[,lw=0.4,bend=-10, style={dashed}](mon1)(pool2)
\Edge[,lw=0.4,bend=-15, style={dashed}](mon1)(fin1)
\Edge[,lw=0.4,style={dashed}](fin1)(pool1)
\Edge[,lw=0.4,bend=1, style={dashed}](mon2)(pool1)
\Edge[,lw=0.4,bend=-35, style={dashed}](mon2)(fin1)
\Edge[,lw=0.5,bend=-10,Direct, style={dashed}](b1)(mon2)
\Edge[,lw=0.5,bend=-10,Direct, style={dashed}](b5)(mon1)
\Text[x=2.7,y=1, width=3cm,fontsize=\scriptsize]{Users}
\end{tikzpicture}
\caption{Centralization in the blockchain ecosystem: mining or staking pools, fin-tech institutions and intermediaries who offer services -- verification, monitoring, data analytics -- on public, permissionless blockchains may add value to the ecosystem but also pose a threat to decentralization. The dashed lines indicate possible interconnections between these entities which may further centralize the system.}\vspace{-0.2cm}
\label{fig:decentralization}
\end{figure}
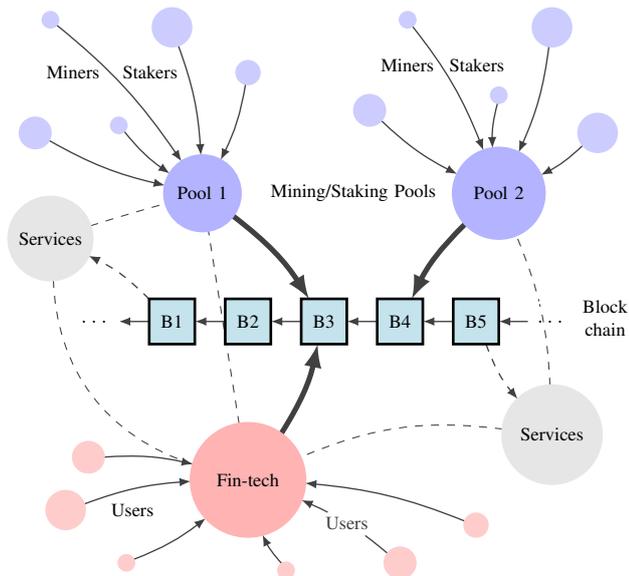

Mining pools are not the only threat to decentralization. Other sources involve the underlying network layer, the geographic or economic motives to concentrate mining rigs in countries with low energy cost, and the increasingly sophisticated technology that is required to participate in the block creation process \cite{Wa18}. \cite{Co17} study anti-trust policies in Turing-complete blockchains, i.e., blockchains that also support smart contract execution, and argue that although smart contracts mitigate information asymmetries and improve social welfare, they also encourage collusions, and hence generate a threat to decentralization. \cite{Leo18} develop a method to bootstrap the blockchain without a genesis block created by a trusted authority. In all cases, the threats of centralization and trust formation raise the closely related question of blockchain governance and sustainability in the long run which is addressed in \Cref{sec:persistence}, \cite{Bo18}. Various sources of centralization in the blockchain ecosystem are illustrated in \Cref{fig:decentralization}.

\subsection{Fairness}
\label{sub:fairness}
An integral element of stability in non-permissioned protocols is \emph{fairness}, which relies on the premise that participating nodes should be rewarded proportionally to their resource contribution, \cite{Chen19}. Recall that each node $n\in \NN$ participates in the protocol by providing some consensus-critical resource $r_n$ which corresponds to a proportion $p_n$ of the total resources. If $U_T$ denotes the total rewards ($U_T$ can be positive or negative) distributed by the protocol to nodes over a (long) period $T$ of time, then we can formally define fairness as follows.
\begin{definition}[\emph{Fairness \cite{Pa17b}}] The \emph{reward allocation mechanism} of a protocol $\Pi$ is said to be $\(\alpha,\epsilon\)$-\emph{fair} for some $\epsilon>0$, if in the presence of an $\alpha$-strong adversary, each node $n\in \NN$ can guarantee $\(1-\epsilon\)\cdot p_n\cdot U_T$ of the rewards over any period of length $T$.
\end{definition}
Achieving fairness seems challenging in practice. Message delays and network latency can cause a disproportional distribution of rewards \cite{Gu18}. Focusing on PoS protocols, \cite{Fa18} introduce the notion of equitability to quantify how much a proposer can amplify her stake compared to her initial investment. Even with everyone following the protocol (i.e., honest behavior), existing methods of allocating block rewards lead to poor equitability, as does the initialization of systems with small stake pools and/or large rewards relative to the stake pool. Consensus in distributed computing with weighted nodes and more general notions of fairness are studied in \cite{Ga11,As16,Wa13}. \cite{Pa17b} extend this notion to environments with adaptive corruption by strengthening the definition of \enquote{ideal protocol quality} defined in \cite{Ga15} and \cite{Pa17a}. \par
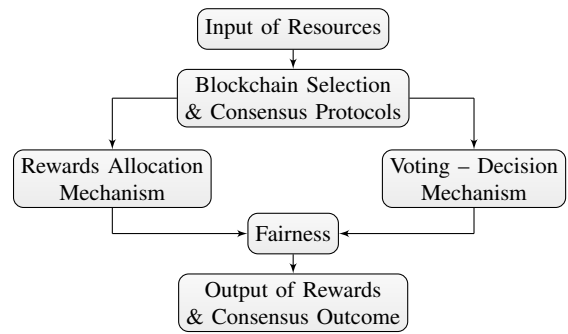
\begin{figure}[!htbp]\vspace{0cm}
\centering
\begin{tikzpicture}[>=latex', scale=0.85, every node/.style={transform shape}]
\tikzset{block/.style={shape=rectangle, scale=1, top color=gray!5, bottom color=gray!15, rounded corners=0.133cm, draw=black, align=center, minimum size=4ex, fill=white}}  

\node [block] (start) {Input of Resources};
\node [block, below=0.3cm of start] (protocol) {Blockchain Selection \\ \& Consensus Protocols};
\node [coordinate, left = 1cm of protocol] (node1){};
\node [coordinate, right = 1cm of protocol] (node2){};
\node [block, below=0.8cm of node1] (rewards) {Rewards Allocation \\ Mechanism};
\node [block, below=0.8cm of node2] (voting) {Voting -- Decision \\ Mechanism};
\node [coordinate, below = 0.4cm of rewards] (node3){};
\node [coordinate, below = 0.4cm of voting] (node4){};
\node [block] at ($(node3)!0.5!(node4)$) (fairness) {Fairness};
\node [block, below=0.3cm of fairness] (output) {Output of Rewards \\ \& Consensus Outcome};

\draw[->] 
(start) edge (protocol) (node1) edge (rewards) (node2) edge (voting) (node3) edge (fairness) 
(node4)-- (fairness)
(fairness) edge (output);
\draw[]
(protocol) -- (node1)  (protocol) -- (node2) (rewards) -- (node3) (voting) -- (node4);
\end{tikzpicture}\vspace{0.2cm}
\caption{Fairness in the two core mechanisms of blockchain protocols: allocation of rewards and aggregation of voting data in the block creation process. Rewards are fairly allocated if they are proportional to participants' resources. However, fairness in proportional voting is a premise that has been theoretically challenged, \cite{Fe98}.}
\label{fig:fairness}\vspace{-0.3cm}
\end{figure}

Fairness in blockchains can be understood as a two dimensional notion that entails both the reward allocation and the block creation mechanism, as illustrated in \Cref{fig:fairness}. Current protocols are based on the premise that proportional voting is fair, \cite{Le19}. However, the simple and seemingly appealing axiom \qt{\emph{one unit of resource (one computer or one coin), one vote}} has been theoretically refuted in traditional voting systems \cite{Ba99,Wo89}.

More importantly, node selection proportionally to their resources -- as in current PoW and PoS protocols -- does not necessarily imply fairness in the voting process, \cite{Fe98}. An illustration is provided in the following example.

\setcounter{example}{2}
\begin{example}
To understand what can go wrong, consider a simplified blockchain network with three nodes that control $p_1=0.45, p_2=0.4$ and $p_3=0.15$ of protocol resources, respectively. The paradox that arises under pure proportional voting, and which is already known in political science, is that the weaker node $3$ is equally powerful with the much stronger node $2$ and in fact more desirable from the perspective of both nodes $1$ and $2$ in forming a majority ($51\%$) coalition.
\end{example}
Interestingly, this example also applies to PoW blockchains that do not involve voting. The reasoning is that any coalition that has a combined majority of protocol resources can control the protocol, thus creating a threat in all permissionless blockchains \cite{Leo19}. 

\section{Efficiency}
\label{sec:efficiency}
After establishing that a blockchain protocol is optimal and stable, the next concern is to efficiently meet these goals in practice. The main question in this context relates to a reasonable use of resources and, in particular, to energy waste, scalability of operations, and benchmarking to centralized solutions. Formally:
\begin{itemize}
\item[Q:] \emph{Does the implemented protocol make efficient use of its resources, and how does it compare to a conventional, centralized solution?}
\end{itemize}
Efficiency of computation, at least from the perspective of time and space, has been thoroughly studied since the 1940s by Turing and von Neumann. Algorithmic game theory and mechanism design study questions on the intersection of optimality, efficiency and stability \cite{Ni07}. While, ideally, a protocol implements a (near) optimal equilibrium efficiently, computational complexity theory implies that there are fundamental limitations of efficient computation \cite{Pa03}. These limitations force us to consider trade-offs, e.g., approximate optimality versus speed or approximation algorithms \cite{Va13}. The different elements of efficiency are discussed below.

\subsection{Energy Consumption}
Efficiency exhibits a tradeoff between a modest (excessive) waste of resources and a high (low) risk of attacks. Nodes in PoW protocols provide proofs of the validity of created blocks via energy consumption, which creates a negative environmental externality, \cite{Kr13}. A promising alternative is offered by the PoS or \emph{Virtual Mining} protocols which reduce this huge energy waste \cite{Ki12,Kw14,Be16,Gi17,Sal17}. Since PoS protocols delegate decision power via proofs of coin (stake) ownership, one of their main advantages over Nakamoto's PoW is their environmental sustainability \cite{Be16}.\par
Energy is not the only input that can be inefficiently used by a protocol. Space for data storage, bandwidth, and Random-Access Memory are only a few \cite{Ch18}. Other aspects of efficiency involve the times to process and finalize transactions and the communication complexity that is required for the distributed network to reach consensus \cite{Za18}. Importantly, different appli\-cations introduce various degrees of uncertainty in the use of such resources and increase the challenge of designing efficient solutions. The processing power used for transaction processing and block propagation delay also determine the outcome of the mining competition \cite{Li18}. Failing to address such issues demotivates agents from participating and leads to centralization. In this sense, efficiency is also related to stability \cite{Di17}. \par
Better ways to utilize the energy spent in PoW protocols may eliminate -- if successful -- the advantage of PoS over PoW protocols in terms of energy waste \cite{Bal17}. \cite{Vu15} provide a classification of other early proposals and open questions in this direction. 
Still, all of these alternative proposals need to tackle the problem of \emph{scalability}, described next.
 
\subsection{Scalability}
Scalability refers to the property that the consensus protocol -- and hence the blockchain -- benefits from the addition of nodes or resources \cite{Zo17}. Generally, a blockchain is scalable if it exhibits \emph{positive scale effects}, i.e., if increased participation leads to (i) increased throughput and (ii) improved liveness, safety, stability and efficiency guarantees. Since these indicators may respond differently to variations in the number of nodes (or the amount of resources), it is more convenient to understand scalability as a property of \emph{performance measures} rather than of the blockchain protocol $\Pi$ as a whole. Recall that a performance measure $U_{\Pi}:\states\to\mathbb R$ is called positive (negative) if increasing values indicate better (worse) performance, cf. \Cref{sec:model}.  
\begin{definition}[Scalability]\label{def:scalability} 
Let state $X$ have consensus-critical resources $r_n$, for nodes $n \in \N$, and state $X'$ resources $r'_n$ such that $\sum_{n \in \N} r_n > \sum_{n \in \N} r'_n$. Then $\Pi$ is scalable in the positive performance measure $U_{\Pi}$ if $U_{\Pi}\(X\)>U_{\Pi}\(X'\)$ for any such $X,X'\in \states$. 
\end{definition}
\Cref{def:scalability} states that a protocol $\Pi$ is scalable in the performance measure $U_{\Pi}$ if an increase in the resources of the current state implies an improved performance for $U_{\Pi}$. The definition for \emph{negative} performance measures is similar. 

\setcounter{example}{0}
\begin{example}[Bitcoin]
To achieve its strong safety guarantees \cite{sirer2016online}, the use of computational resources by the Bitcoin (PoW) protocol is not efficient: the maximum transaction throughput is the same as five years ago despite a dramatic increase in hash rate and energy consumption \cite{Od14}. \cite{Bi17} identify excessive spending and inefficiencies in the prevailing equilibrium of Bitcoin's \emph{follow-the-protocol} strategy. \cite{Chi17} suggest that partial or complete substitution of energy-costly mining activities with PoS mechanisms could benefit Bitcoin and make it more efficient in the long run. \par
Attacks on Bitcoin can inflict a significant energy cost on miners \cite{Lu15b}. In general, by partitioning the network or by either censoring or delaying the propagation of blocks, network-layer attacks can cause a significant amount of mining power to be wasted, leading to revenue losses and enabling a wide range of attacks such as double-spending. To deal with these threats, \cite{Lu17} propose a mining pool that will run as a smart contract and show that this is a solution with good efficiency and scaling properties.
\end{example}

Currently, a broadly studied solution to scalability is \emph{sharding}, see e.g., Elastico \cite{Luu16}, OmniLedger \cite{Ko18} and \mbox{Ethereum 2.0} \cite{Bu18online7,Et18online2}. As an alternative approach, \cite{Gaz18} model the concept of sidechains as a means to increase scalability and enable the interoperability of blockchains. Their construction features merged-staking which prevents \emph{Goldfinger attacks} -- attacks whose explicit goal is to undermine and destabilize the consensus protocol \cite{Kr13,Bo18} -- and cross-chain certification based on novel cryptographic primitives. \cite{Ba17a} study a similar combination of consensus protocols with PoS subchains linked to the PoW Bitcoin blockchain.\par

\subsection{Throughput}

Although throughput is closely related to scalability, a protocol can prioritize throughput even without making the protocol fundamentally more scalable. For example, by increasing the maximum number of transactions per block (e.g., Bitcoin Cash vis-\`a-vis Bitcoin), throughput is increased without essentially affecting scalability. The same is true for protocols such as EOS.IO and TRON, which achieve much higher throughput than, e.g., Bitcoin and Ethereum by curtailing the number of potential block proposers. In fact, a BFT protocol can easily achieve much higher throughput than a Nakamoto protocol if the number of nodes, denoted by $N$, is low. However, such protocols typically suffer from \emph{negative} rather than positive scale effects when the number of nodes increases due to the ${\O}(N^2)$ message complexity. So it is possible for a protocol change to have a positive effect on throughput yet a negative effect on scalability. This informs our definition below.

\begin{definition}[Throughput]\label{def:throughput} 
Let the performance measure $U_{\Pi}^*(X)$ denote the long-term average transaction throughput, starting from state $X\in \states$. A blockchain protocol $\Pi$ with resources $r_n$, ${n \in \N}$ has a higher throughput than another protocol $\Pi'$ with the \emph{same} resources if $U^*_{\Pi}(X) > U^*_{\Pi'}(X)$.
\end{definition}

Fundamentally, scalability concerns the effects on the outputs when the resources are changed and the protocol is kept the same, whereas throughput (as a PREStO category) concerns the effects on the outputs when the protocol is changed and resources are kept the same.
 
\subsection{Centralized Systems as Benchmarks} 
From a managerial perspective, the integral question in launching a blockchain project or application is whether a blockchain is indeed better than a centralized system for the intended purpose \cite{wust2018you,Del19online}. 
Since blockchains eliminate trusted authorities to reach consensus via the coordination of distributed and self-interested entities, several questions come into play. How does the distributed system compare to a benchmark solution? Does it provide improved performance in terms of costs, efficiency and security? \par
Interestingly, related questions have been thoroughly researched in game theory. In particular, traffic routing, queueing theory and congestion networks explore precisely these tensions between equilibration and efficiency of centrally regulated systems, \cite{Ch05}. The trade-offs are quantified by the \emph{Price of Anarchy} (PoA), which measures the sub-optimality caused by self-interested behavior relative to centrally designed and socially optimal outcomes \cite{Ro09,Im10,Pi16}. PoA is defined as the ratio between the performance of the system at the worst-case equilibrium and that at a socially optimal state \cite{Ko99}. 
\par
Studying this question in the current context requires us to quantify different aspects of blockchain performance and compare them to either an existing or a socially optimally (ideal) solution provided by a benevolent social planner or authority. To measure the effects of \emph{decentralizing} a system when implementing it as a blockchain, we evaluate a derivative notion, the \emph{Price of Decentralization} (PoD), which can be defined as
\begin{equation}\label{eq:pod}\pod\(U_\Pi\):=U_\Pi\(D,n\)/U_{\Pi}\(D,1\)\end{equation}
As above, $U_\Pi:\Pi \to \R$ denotes a performance measure of protocol $\Pi$. PoD compares the performance of the blockchain at state $\Pi$ in which the system is operated by $n$ nodes who all follow the protocol, i.e., use strategy profile $D=\times_{n\in\NN}d$, to its performance when it is operated by a single node \emph{and} in an optimal way. PoD retains the flavor of PoA but isolates the effect of decentralization. We emphasize the following:
\begin{itemize}[leftmargin=*]
\item Depending on the application, the denominator may refer to an optimal solution provided by an ideal social planner, or a centralized solution provided by a revenue-maximizing intermediary, or both. 
\item The numerator evaluates the system's performance under the assumption that \emph{all nodes follow the protocol}. Thus, it does not account for strategic behavior; instead it assumes a default decentralized protocol execution and compares it to the corresponding centralized solution. More relevant in this direction is the approach of \cite{Mo09} who introduce the price of malice, to study how a system degrades in the presence of malicious agents.
\item Unlike PoA, the range of values of PoD is not restricted by $1$ and depends (i) on whether $U$ is a \emph{positive} or \emph{negative} performance measure, cf. \Cref{sec:model} and (ii) on whether the blockchain solution improves over the performance of the centralized solution (although this is unlikely) or not.
\end{itemize}

\setcounter{example}{3}
\begin{example} To illustrate the above, let $U_\Pi:\Pi\to\N$ be the number of messages that need to be exchanged between $N$ nodes to reach consensus according to protocol $\Pi$. In a fully centralized execution of the system, the single entity trivially needs to send one message to each node to inform them of the the decision, leading to ${\O}(N)$ messages in total. However, in a BFT protocol, in which every node has to send a message to each other node, consensus takes $\O\(N^2\)$ messages, see e.g., Solidus, Algorand or Elastico, \cite{Ba17}. In this case, $\pod\(U_\Pi\)=U_\Pi\(N\)/U_\Pi\(1\)= \O\(N^2\)/\O\(N\)= \O\(N\)$. This shows that the PoD of the BFT protocol $\Pi$ concerning the communication complexity $U_\Pi$ is linear in the number of nodes $N$ and hence, is unbounded as $N$ grows to infinity.
\end{example}

\section{Robustness}
\label{sec:robustness}
Suppose that a protocol has provable performance guarantees within its scope (optimality) and that the \emph{follow-the-protocol} strategy is an equilibrium (stability), at which the protocol resources are reasonably utilized (efficiency). The next natural step in protocol design is to explore how smoothly and rapidly the protocol's properties degrade when we move away from the vanilla setting. These concerns are expressed by the following question:
\begin{itemize}
\item [Q:] \emph{What is the resistance of the protocol to perturbations on its underlying assumptions?} 
\end{itemize}
In the case of a parametrizable protocol, this question may also be phrased in terms of the extent of the parameter variation that the system can tolerate \cite{As10}. Essentially, robustness tests the assumptions that were used to equilibrate and stabilize the system. The main challenge is to assess protocol performance under conditions that are not captured by the \emph{ideal} setting of \Cref{def:Nash}, such as parameter fluctuations, collusion between nodes, and malicious or irrational behavior \cite{Hu18}.

\subsection{Alternative Equilibrium Concepts}
The application of the Nash equilibrium as a stability concept in blockchains is not entirely uncontroversial \cite{Ha11, Az18}. In particular, \cite{Ha11} and \cite{Da06} discuss the following shortcomings of Nash equilibria in distributed computational systems: unexpected behavior (irrational players with out-of-system incentives), coalitional deviations, computational limitations (resource-bounded players), and too much uncertainty or a lack of information (players are unaware of all the aspects of the game). To deal with these issues, \cite{Ab06,Ha11} propose the notion of \emph{robust strategy profiles} which have two defining components. On one hand is the profit of deviating players. If $k$ agents simultaneously deviate from a given strategy profile but are not able to increase their profits, then the strategy profile is said to be \emph{$k$-resilient}. On the other hand is the harm incurred to non-deviating players. If $t$ agents simultaneously deviate from a given strategy profile but are not able to decrease the profits of non-deviating agents, then the strategy profile is said to be \emph{$t$-immune}. Combining these two elements yields the notion of $\(k,t\)$-robust equilibria as strategy profiles that are both $k$-resilient and $t$-immune.\par
Despite its theoretical appeal, \cite{Ha11} observes that the concept of $\(k,t\)$-robust equlibria has its own limitations and points to concepts of computational equilibria and particularly to the BAR-model -- model with Byzantine, Altruistic and Rational agents -- as possible alternatives \cite{Ai05,Az18}. Nevertheless, \cite{Gr14} provide strong arguments to support the use of Nash equilibria by showing that large games are innately fault-tolerant. In fact, \emph{anonymous games} that can be used to model blockchain mining are shown to be resilient against irrational behavior (Byzantine faults), coalitions and asynchronous play.\par
In an approach that is particularly relevant to the blockchain setting, \cite{Li16} define robustness of an equilibrium as the maximum proportion of malicious nodes that the desired equilibrium strategy can tolerate, in the sense that this strategy remains the best strategy for rational players. In this definition, robustness is understood as a \emph{local} property, i.e., as a property of a specific strategy profile and against specific adversarial strategies. This definition overcomes the computational difficulties of defining robustness for the blockchain protocol as a whole, and utilizes the fact that in blockchains, the analysis of robustness mainly concerns the \emph{follow-the-protocol} strategy.

\subsection{Out-of-Protocol Incentives}

In reality, an adversarial node may try to change the behavior of other nodes by influencing their utility functions through threats or rewards. One of the earliest examples of this is \emph{feather forking} \cite{Mi13online} in Bitcoin: in this case, a miner threatens to refuse to extend blocks if they contain a blacklisted transaction. Even if the expected impact of the threat is small, it may be high enough compared to the small cost of enforcing the blacklist to make it rational to comply with the threat. Similarly, \emph{bribery} \cite{Bo16,Bo18,mccorry2018smart} or \emph{discouragement} \cite{Bu18online8} attacks can be used to distort the incentives of rational nodes. 

In protocols in which it is known how much consensus-critical resources are owned by each of the nodes (i.e., semi-permissionless or permissioned blockchains), it may be possible to predict which nodes are scheduled to propose blocks in the near future. Accordingly, \cite{Br18} identify two complementary properties -- \emph{recency} and \emph{predictability} -- of all longest-chain PoS protocols and devise relevant attacks to show that all such protocols are susceptible to certain kinds of malicious behavior. Finally, \cite{Se18} explore the trade-offs between PoW and PoS consensus and find that a combination of both may yield robust results. In particular, for small numbers of participants PoS exhibits better security properties against 51\% attacks by mining pools but as the size of the network increases, they recommend reverting to PoW. 

\subsection{Resistance to Malicious Behaviour}

Not all nodes are solely interested in protocol rewards -- for example, they may be interested in performing a \emph{Goldfinger attack} \cite{Kr13}, in which one cryptocurrency platform is attacked to increase the value of others. One way of modeling this is to assign to such an attacker a utility that is the inverse of the collective utility, and calculate the total losses under the new equilibrium. Another approach is to calculate bounds on the losses that attackers can do relative to their own losses. In \cite{Bu17online,Bu18online8} this is made explicit through the \emph{griefing factor} (GF). In particular, let $n,m \in \N$ be nodes, $x_0 \in \states$ a starting state, $D$ the strategy profile where all nodes play the default strategy, and $D_{n,s}$ the profile where all nodes play $d$ except $n$ who plays $s$. Then the griefing factor of between $n$ and $m$ is defined as
$$
\text{GF}(n,m,s) = \frac{v_{m}(D_{n,s},x_0) - v_{m}(D,x_0)}{v_{n}(D_{n,s},x_0) - v_{n}(D,x_0)}
$$
if the denominator is positive, and $\infty$ otherwise. Summing over all $m$, yields the \emph{network's griefing factor} (NGF). The value of the NGF reflects the relative loss that a participating node needs to incur in order to inflict a ``unit'' of loss on the whole network. For instance, a value of NGF equal to $1$ implies that node $n$ needs to destroy $1$ dollar for every $1$ dollar loss that she inflicts to the network. Accordingly, larger values of the NGF correspond to the most harmful cases. 

\section{Persistence}
\label{sec:persistence}
The four PREStO categories discussed so far consider the protocol when it operates \emph{at} or \emph{near to} equilibrium conditions. However, what happens if the protocol is forced away from its equilibrium, for instance after a large-scale attack or catastrophic black swan event? Hence, to establish a protocol's persistence property, we ask the following question:
\begin{itemize}
\item[Q:] \emph{Does the protocol have mechanisms to recover from highly non-equilibrium conditions and return to stability in its optimal state? If so, then how fast, and at what cost?}
\end{itemize}
These questions deal with the long-term sustainability of the blockchain platform. Whereas for robustness, we studied performance under perturbations of the stability assumptions, for persistence, we take this idea to its logical extreme. We assume that the ecosystem is under a large-scale or protracted attack, and study whether it is designed to recover and resume its desirable properties, at least sufficiently often. Hence, we want to assess to what extent a blockchain has the qualities to survive and evolve under extreme crashes, technology shocks or other rare events.

\subsection{Weak \& Strong Persistence}
To understand protocols from this perspective, we formalize the notions of \emph{weakly} and \emph{strongly} persistent properties in the blockchain context. These ideas have been introduced within evolutionary game theory and in the study of biological systems, i.e., recovery of an ecosystem after infection from a virus \cite{Ho98,Sm11}. More relevant to the current context is the combination of these ideas with tools from optimization theory and algorithm design \cite{Be01,Pi14a}. Formally, recall that a property or \emph{feature} of a protocol $\Pi$ is defined as a function $\F_\Pi:\states\to\{0,1\}$, cf. \Cref{sec:model}.
\begin{definition}[\emph{Weakly \& strongly persistent properties} \cite{Pi14b}] Consider a protocol $\Pi$ and a property $\F_\Pi: \states \to \{0,1\}$. Let $\(X_t\)_{t\geq0}$ be a protocol execution with initial state  $x_0 \in \states$. We say that $\F_\Pi$ is 
\begin{itemize}[topsep=0pt, noitemsep, align=left, leftmargin=*]
\item \emph{weakly persistent} for protocol $\Pi$, if for any $x_0$, and any $T>0$, there exists $t>T$ such that $\F_\Pi\(X_t\)=1$.
\item \emph{strongly persistent} for protocol $\Pi$, if for any $x_0$, there exists $T>0$, such that $\F_\Pi\(X_t\)=1$ for all $t>T$. 
\end{itemize}
\end{definition}
Intuitively, a \emph{weakly persistent} property will eventually be satisfied and become satisfied again infinitely often given any initial system condition, whereas a \emph{strongly persistent} property will eventually be satisfied and \emph{stay} satisfied given any initial system condition, \cite{Pi14b}. These definitions capture the idea that a desirable property may not be satisfied by a system in equilibrium, but in a dynamic way. This allows for more flexibility between recovery/convergence time, \qt{periodicity}, and the cost of implementation. \par

\setcounter{example}{4}
\begin{example}[The Blockchain Trilemma] The idea of supporting two or more incompatible properties in a weakly persistent manner, as described above, can be exploited to address the challenging \qt{Blockchain Trilemma} (\cite{Ab18,Co19}) which is illustrated in \Cref{fig:trilemma} and which we will discuss in more detail in \Cref{sec:trilemma}. The vertices of the triangle correspond to the three seemingly incompatible but desirable properties that blockchain consensus protocols may satisfy: \emph{decentralization}, \emph{scalability} and \emph{safety}. Protocols can be thought of as points inside the triangle, with coordinates indicating the degree of satisfaction of each of these properties. \par
Designing an optimal protocol -- i.e., a protocol which in equilibrium satisfies simultaneously all three properties -- has been a formidable task for blockchain architects, \cite{Gi17}. Such a protocol is indicated by the green dot in \Cref{fig:trilemma}. However, the idea of weak persistence can be exploited for an alternative design: a protocol could solve the trilemma by constantly alternating between states that satisfy a non-conflicting subset of the otherwise incompatible properties. This is captured by the blue dot protocol and the dashed arrows in \Cref{fig:trilemma} which show its transition between different states.
\begin{figure}[!htbp]
\vspace{-0.2cm}\centering
\begin{tikzpicture}[>=latex',node distance = 4cm]
\node (S2) {Decentralization};
\node [left of = S2] (S1) {Scalability};
\node [above of =S2, node distance=3cm] (Phant) {};
\node [left of = Phant, node distance =2cm] (S3) {Safety};
\draw (S1) -- (S2) -- (S3) -- (S1);
\node at (-2,1)[circle,fill,green, inner sep=1.5pt]{};
\draw [->,dashed,blue] (S2) to [bend left=-20] node[circle, fill,blue, inner sep=1.5pt] {} (S1);
\draw [->,dashed,blue] (S1) to [bend left=-20] node[circle, fill,blue, inner sep=1.5pt] {} (S3);
\draw [->,dashed,blue] (S3) to [bend left=-20] node[circle, fill,blue, inner sep=1.5pt] {} (S2);
\end{tikzpicture}
\caption{Dealing with the Blockchain Trilemma: The green dot denotes an ideal protocol that satisfies all three properties in equilibrium. The blue dot denotes a protocol that cycles around the ideal solution and which satisfies the incompatible properties in a weakly persistent (recurrent) manner.}\vspace{-0.3cm}
\label{fig:trilemma}
\end{figure}
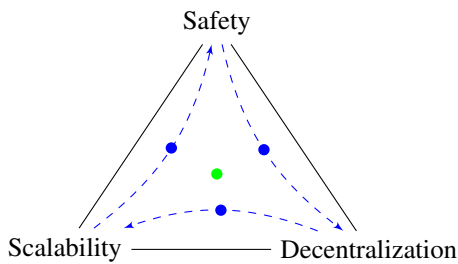
\end{example}
The idea of studying distributed computation through the lens of dynamical systems has been recently inititated by \cite{Ja17}. Based on their ideas, persistence can be also used to formulate a weaker definition of fairness, cf. \Cref{sub:fairness}. Namely, a protocol can be described as \emph{fair} if each node gets to be selected in the block creation process infinitely often. 

\subsection{Recovery from Majority Attacks} 
\label{sec:recovery}
One of the major challenges in blockchain consensus protocols is the recovery from attacks by malicious nodes who control the majority of protocol resources, \cite{Bo18}. Existing protocols establish their safety and liveness properties under the assumption of either a simple -- $51\%$ -- or an enhanced -- usually $67\%$ -- honest majority of nodes, \cite{Zh18}. Contrary to initial beliefs that these attacks are only of theoretical interests, recent studies have documented the contrary, \cite{Bo16}. An important insight from these studies is that it is sufficient to gain control for some short period of time, for instance by temporarily renting protocol resources. 

A suggested mechanism to recover from majority or large-scale attacks on the Ethereum blockchain is the \emph{minority fork}, proposed by \cite{Bu18online7}. In brief, a minority fork is a mechanism to recover the majority of the consensus-critical resources through a fork initiated by an honest minority. Because the majority cannot create blocks on both branches of the fork, they will be seen as offline on the minority-initiated branch, which may cause their share to shrink on this branch. Such a scheme is fundamentally impossible in permissionless blockchains. 

\subsection{Governance \& Sustainability}
\label{sub:governance}
Persistence is closely related to the decision processes that determine the structure and operation of the blockchain. The practical need for an optimal governance structure in the Bitcoin community has already been observed by \cite{Kr13}. In a different approach, \cite{Cat16} view the blockchain as a public good and discuss the role of intermediaries that will provide paid services of blockchain verification and monitoring which adds value to the entire blockchain ecosystem. With the exception of some tentative predictions, the formal governance structure of public, permissionless blockchains has yet to be determined, \cite{Ho19online}. The issues of governance and long-term sustainability in blockchains are integral to their success, and therefore central themes in their evolution. 

\section{Evaluation: Use Cases of the PREStO Framework}
\label{sec:evaluation}
\begin{figure*}[!t]
\vspace{-0.2cm}\centering
\includegraphics[clip, scale=0.61]{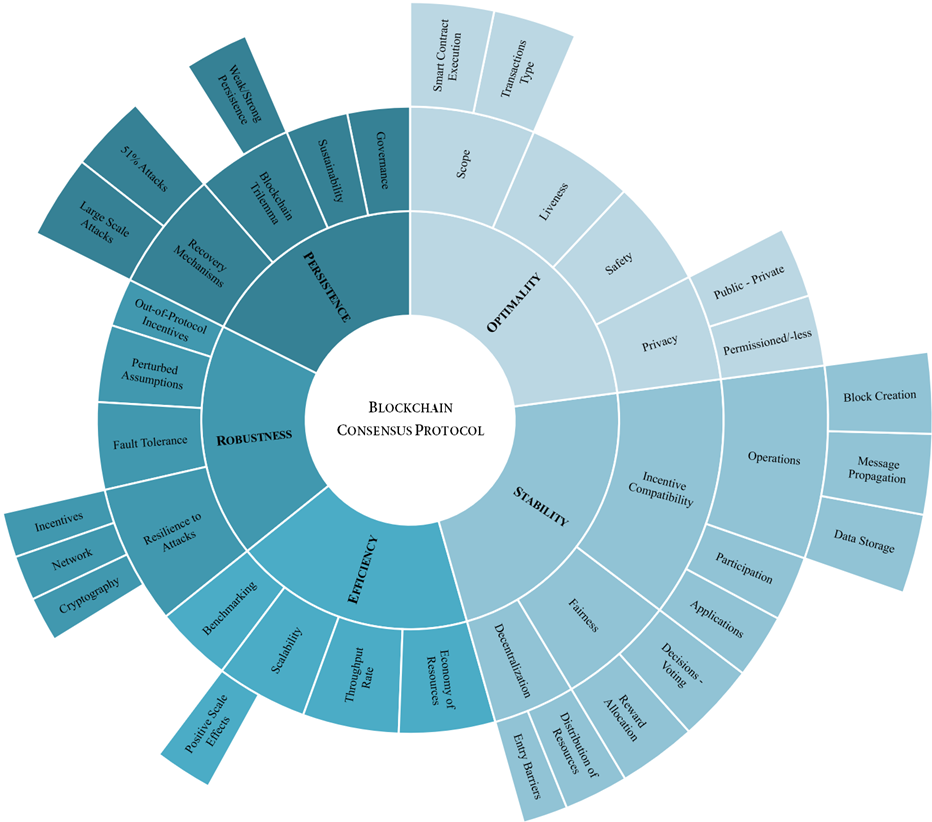}
\caption{Visual representation of the PREStO framework. The Blockchain consensus protocol lies in the middle of a series of concentric circles. The inner cycle comprises the 5 major axes of PREStO and the outer cycles correspond to subcategories of increasing granularity. Starting from this setting, the framework can be extended in a dynamic way to integrate features of more elaborate blockchains in the future.}
\label{fig:presto_detailed}\vspace{-0.2cm}
\end{figure*}

In this section, we evaluate the PREStO framework's ability to illustrate the fundamental differences between various protocols. In Section~\ref{sec:trilemma}, we begin by comparing the PREStO framework to the Blockchain Trilemma, another well-known model of protocol properties. Next, we use the PREStO framework to illustrate the properties of a range of $9$ recently proposed protocols and protocol modifications. As can be seen from our analysis, the PREStO framework reveals more detail than the blockchain trilemma. We conclude the section with an overview of research challenges in \Cref{sec:challenges}. Throughout the next section we refer to the visual summary of the PREStO displayed in \Cref{fig:presto_detailed}. 

\subsection{The Blockchain Trilemma}
\label{sec:trilemma}
In \Cref{sec:persistence}, we referred to the Blockchain Trilemma as an example of how the theory of dynamical systems can offer a different perspective to one of the most long-standing problems in distributed computing. In particular, largely incompatible properties, like safety, decentralization and scalability, can be satisfied in a weakly or strongly persistent manner, i.e., in a regularly alternating but certainly recurrent manner. This offers a potential way out of the current deadlock in the search for an ideal protocol that will concurrently satisfy all of the three highly desirable properties.\par
But how do the properties -- safety, decentralization and scalability -- of the Blockchain Trilemma relate to the PREStO framework? Or put differently, does PREStO provide the right tool to parse the trilemma in its basic components, to communicate it to experts of different backgrounds and ultimately to reason about and try to resolve it? Before we can answer these questions, we discuss the three properties of the blockchain trilemma in more detail. We base our description on \cite{Et20online}, which gives a concise, formal description of the three properties. In the following, we assume for simplicity that there is a single consensus-critical resource. We also introduce the scale parameter $\nscale$ which represents the \qt{size} of the network. The properties of the blockchain trilemma describe the behavior of the blockchain as it grows in size, i.e., in the limit $\nscale \rightarrow \infty$. In the following, the total number of nodes in the network is given by $\nscale$, and $R\(\nscale\)$, the total amount of resources in the network, is proportional to $\nscale$ such that $\lim_{\nscale\rightarrow \infty} R\(\nscale\)/\nscale = c$ for some constant $c > 0$. The three properties are then as follows.
\begin{description}[noitemsep,leftmargin=*]
\item[Safety:] The network is \qt{secure} against an attacker whose resources are proportional to $\nscale$. To make this formal: let the attacker's resources be denoted by $r_a(\nscale)$ when the network size is $\nscale$. The attacker's resources are proportional to the network size if $\lim_{\nscale\rightarrow \infty} r_a(\nscale)/ \nscale = \alpha$ for some constant \mbox{$\alpha > 0$}. The system must still be secure in this setting.
\item[Decentralization:] An honest node whose resources are proportional to $c$ but not $\nscale$ is still able to \qt{participate} in the network. That is, if the honest node's resources are given by $r_{h}(\nscale)$, then even if $\lim_{\nscale \rightarrow \infty} r_{h}(\nscale) / \nscale = 0$, then the node should still be able to participate.
\item[Scalability:] The system is able to process a number of transactions that is proportional to $\nscale$ rather than $c$. That is, if the number of processed transactions in a network of size $\nscale$ is given by $U^*_{\Pi}(\nscale)$, then we require that $\lim_{\nscale \rightarrow \infty} U^*_{\Pi}(\nscale) / \nscale = \beta$ for some $\beta \in (0, \infty]$, and that  $\lim_{\nscale \rightarrow \infty} U^*_{\Pi}(\nscale) / c = \infty$.
\end{description}
As we can see, the blockchain trilemma describes an asymptotic regime in which certain properties must hold, but leaves these properties open to interpretation. In particular, \qt{security} in the trilemma's safety property can either refer to liveness or safety of \emph{optimality} as displayed in \Cref{fig:presto_detailed}. However, a deeper analysis might also consider bribery or discouragement attacks, which are subcategories of \emph{robustness}. The notion of \qt{participating} in the protocol under the trilemma's decentralization property can refer to the ability to perform the full range of operations, to the existence of entry barriers, or to the ability to be fairly rewarded. All of these are subcategories of the \emph{stability}. Finally, the trilemma's notion of scalability is clearly related to its namesake in PREStO, which is a subcategory of \emph{efficiency}. However, unlike the trilemma, which requires that throughput is at least linear in term of the network size, we merely require that it is increasing (one extension of \Cref{fig:presto_detailed} would be to expand the scalability category to include more subcategories for different rates of growth).

\begin{figure}[!t]
\vspace{-0.2cm}\centering
\includegraphics[clip,  trim=0cm 0.7cm 0cm 0.7cm, scale=0.3]{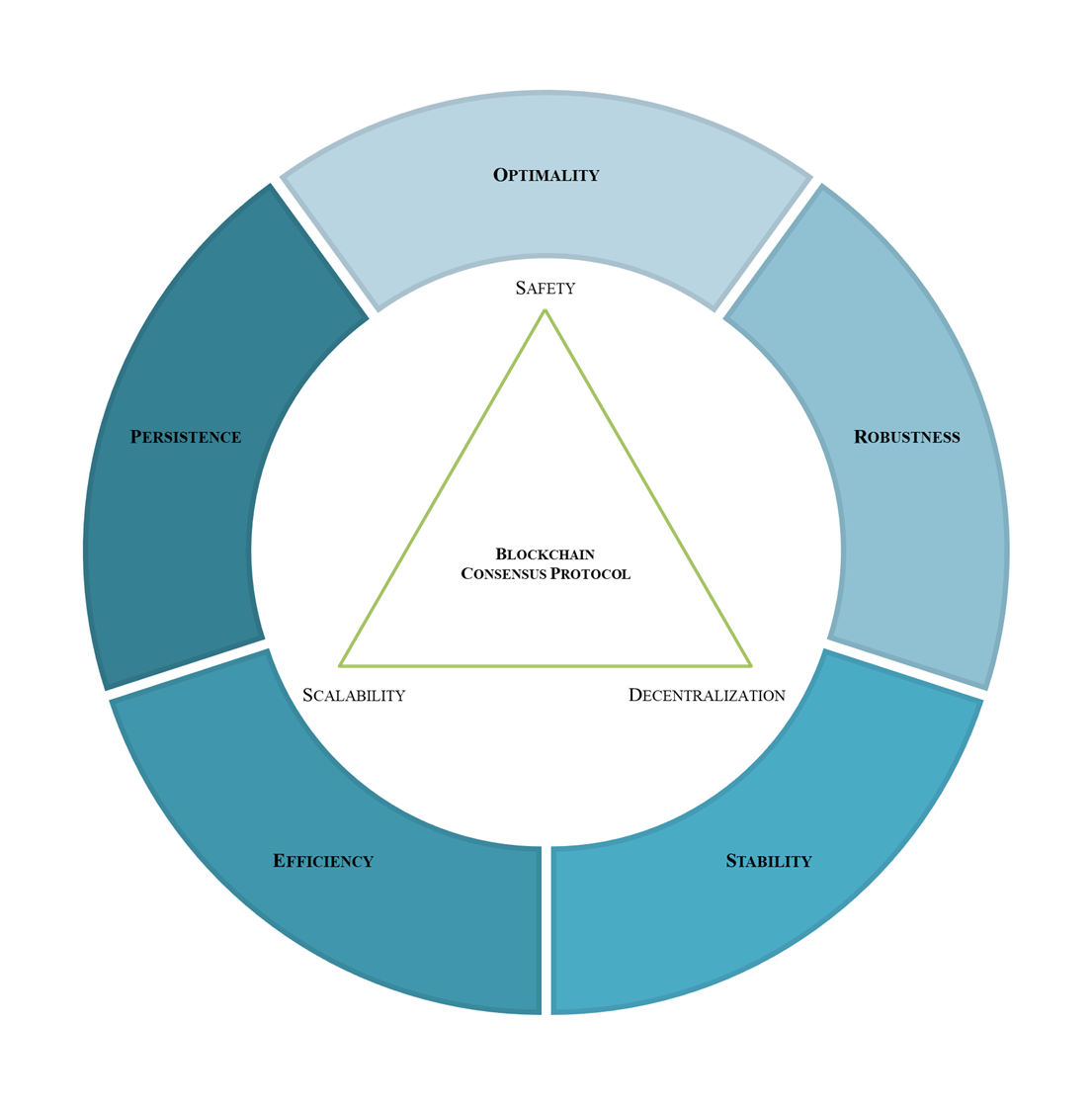}
\vspace{-0.2cm}
\caption{Visual representation of the Blockchain Trilemma in relation to the PREStO framework. The trade-off between Safety, Scalability and Decentralization is precisely captured by the corresponding subcategories in Optimality, Efficiency and Stability. Robustness and Persistence offer alternative approaches for a long-term resolution of the Trilemma.}
\label{fig:trilemma_2}
\end{figure}

Hence, the PREStO framework can be used to fill the gaps and ambiguities left in the definition of the blockchain trilemma. In general, safety relates to optimality (and perhaps robustness), decentralization to stability, and scalability to efficiency. This relationship is illustrated in \Cref{fig:trilemma_2}. 

Turning to practical examples, Bitcoin is considered to be a safe blockchain because an attacker needs to control at least a given fraction of the total network hashrate for most attack to become possible or profitable (e.g., approximately one third for selfish mining). However, Bitcoin does not scale in its current form, and the emergence of pool-mining has lead to an undesirably high centralization, \cite{Arn18,Leo19} -- i.e., it is practically impossible for smaller nodes to perform certain protocol operations such as block creation. There is also a trade-off between decentralization and safety: while solo mining outside of pools leads to higher degree of decentralization, the large variance of mining rewards may force solo miners to drop out of the mining process due to temporal losses and hence, ultimately decrease the network safety, \cite{Pra19}.\par
On the other hand, EOS.IO, \cite{Eos19}, promises both decentralization and scalability. However, although its minor nodes are able to vote, the ability to create blocks is concentrated in the hands of a small number of nodes. As such, it is unclear if EOS.IO can be described as decentralized. Its relatively small number of active nodes also constitutes a barrier to understanding its exact limits in terms of safety in a large-scale public setting \cite{Pra19}. Similarly, Litecoin and Bitcoin Cash \cite{Lit19,Bit19} aim to increase scalability -- via an increased block frequency or size -- 
but although this increases throughput by a constant factor, the network is still not scalable in the sense that throughput is linear (or even increasing) in terms of the network size. Hyperledger, \cite{Hyp19} aims to provide both scalability and safety by operating as a private blockchain, i.e., in a less decentralized setting. Algorand, \cite{Gi17} and Ethereum, \cite{Eth19}, are actively researching or developing ideas like Proof of Stake, sharding, side chains and more efficient Byzantine Fault Tolerant mechanisms that will lead to revolutionary solutions of the Blockchain Trilemma. The list of approaches does not end here, with protocols such as Zilliqa, \cite{zilliqawp}, offering yet more ideas to this debate by using sharding to create a scalable but still decentralized protocol. Finally, while many claim that solving the trilemma is essentially not possible, co-existence of multiple, interoperable blockchains may be another approach to go past this bottleneck in the future, \cite{Med19}. 

\subsection{Evaluating Features of Blockchain Protocols}
\label{sec:properties}

\begin{table*}[!hbt]
\centering
\begin{tabular}{ll ll !{\color{scolor}\vrule}ll!{\color{scolor}\vrule} ll!{\color{scolor}\vrule} l!{\color{scolor}\vrule}l } 
Feature & Example Protocol & \trot{Transaction Scope} & \trot{Privacy} & \trot{Decentralization} & \trot{Fairness} &  \trot{Scalability} & \trot{Throughput} & \trot{Attack resistance} & \trot{Recoverability} \\
 && \multicolumn{2}{c}{\cellcolor{ocolor}} & \multicolumn{2}{c}{\cellcolor{scolor}} & \multicolumn{2}{c}{\cellcolor{ecolor}} & \multicolumn{1}{c}{\cellcolor{rcolor}} & \multicolumn{1}{c}{\cellcolor{pcolor}}\\\\[-0.3cm]
\bottomline\\[-0.25cm]
Partial solutions & \emph{FruitChains} \cite{Pa17b}, \emph{StrongChain} \cite{szalachowski2019strongchain} &  & & \PLUS & \PLUS & & \MINUS  & \PLUS & \\[0.05cm]
Smart contracts & \emph{Ethereum} \cite{Bu14online} & \PLUS & & & & \MINUS & \MINUS & & \\[0.05cm]
Checkpointing & \emph{Casper FFG} \cite{buterin2017casper,Bu19} &  &  & &  &  & \MINUS &  \PLUS & \PLUS \\[0.05cm]
Weighted Voting &\cite{Le19} & & & & & & \PLUS & \PLUS & \\[0.05cm]
Zero-knowledge proofs & \emph{Zcash}& & \PLUS & & & & \MINUS & & \\[0.05cm]
Increased block size & \emph{Bitcoin Cash} &  &  & \MINUS & \MINUS  & & \PLUS & & \\[0.05cm]
Increased block frequency & \emph{LiteCoin} &  & & \MINUS & \MINUS &  & \PLUS  & & \\[0.05cm]
Microblocks & \emph{Bitcoin-NG} \cite{eyal2016bitcoin} & & & \MINUS & \PLUS &  &  \PLUS &  & \\[0.05cm]
Sharding & \emph{Zilliqa} \cite{zilliqa}, \emph{OmniLedger} \cite{Ko18} & &  & & & \PLUS & \PLUS & \MINUS & \\\\[-0.25cm]
\bottomline
\end{tabular}
\vspace*{0.2cm}
\caption{Overview of the impact of several illustrative protocol features on the PREStO categories.}
\label{tab:two}
\end{table*}

In this section, we use PREStO framework as a tool to compare and evaluate a range of recent protocol modifications. A summary of this comparison is presented in \Cref{tab:two}, and we present a more detailed discussion in the following.

\begin{description}[noitemsep,leftmargin=*]
\item[Partial solutions:] Several recent works \cite{Pa17b,szalachowski2019strongchain} have proposed to modify Nakamoto consensus by allowing miners to include unsuccessful attempts to solve the PoW puzzle -- \emph{partial solutions} -- in the blocks. These partial solutions contribute to the block's likelihood to be selected by the fork-choice rule, and award rewards to the finders. Hence, weaker miners are rewarded more often, which reduces their barriers to entry and as such increases \emph{decentralization}. In addition, the inherent advantage that big miners have when confronted with network latency is reduced, which improves \emph{fairness}. Finally, preliminary experiments suggest that it is harder for selfish miners or attackers with a minority of the hash power to overturn blocks, which helps \emph{incentive compatibility} and \emph{attack resistance}. The downside is that new data is added to the blocks, which consumes bandwidth and therefore harms \emph{throughput}. 
\item[Smart contracts:] The main innovation of the Ethereum platform \cite{Bu14online} was to extend the functionality of blockchain from token transfers to the creation and execution of Turing-complete programs called \emph{smart contracts}. As such, the \emph{transaction scope} is much wider. This comes at a cost to \emph{throughput} as nodes need to expend considerably more processing power to execute the transactions and update the global state. Finally, smart contract platforms have a more complicated global state than those with just token transfers -- this has an impact on the applicability of certain sharding techniques (e.g., transaction vs.\ state sharding) \cite{Jia2018,Et20online} and therefore potentially reduced \emph{scalability}. 
\item[Checkpointing:] Casper the Friendly Finality Gadget (FFG) \cite{buterin2017casper,Bu19}, a checkpointing protocol for Ethereum, introduces a formal scheme for nodes to create finalized blocks that cannot be overturned without a manual reset. This increases the \emph{attack resistance} of protocols. Furthermore, Casper FFG enables  the \qt{minority fork} mechanism described in Section~\ref{sec:recovery}, which increase the \emph{recoverability} from majority attacks. However, the voting mechanism that is used for finalization consumes bandwidth, and hence reduces \emph{throughput}.
\item[Weighted voting:] In \cite{Le19}, a modification to PoS protocols was proposed that weighted the consensus power of nodes not just by their staked deposits, but also by their voting history. For example, if they regularly fail to vote for the blocks on the main chain in a timely manner, then their voting power is reduced. This increases \emph{throughput} because offline nodes are less likely to be elected as block proposers, which means that less time is wasted. By contrast, no new data is added to the blocks and the processing power required for clients to update the voting profiles is negligible. Furthermore, \emph{attack resistance} is improved as attackers who seek to harm liveness by not voting (correctly) rapidly lose consensus power.
\item[Zero-knowledge proofs:] In Bitcoin, tokens are protected through a requirement that any token transfer (through the spending of UTXOs) can only be done if correct signatures are produced for the addresses of the sent tokens. These addresses (and the values of individual UTXOs) can be masked using zero-knowledge proofs, which have been implemented in the Zcash platform. This improves the \emph{privacy} of users, but at the cost of additional computing power required to process transactions.
\item[Increased block size/frequency:] By increasing either the size or frequency of blocks (as done in Bitcoin Cash and LiteCoin, which are forks/spin-offs of Bitcoin), the \emph{throughput} in terms of the maximum number of  transactions per second can be increased. However, this also requires that more resources (especially bandwidth) are consumed per second. Since it is more difficulty more smaller parties to handle this additional load, the barriers to entry are increased which leads to worse \emph{decentralization}. Furthermore, the impact of network latency is increased, and since larger nodes have an advantage in high-latency situations, \emph{fairness} is reduced. 
\item[Microblocks:] The Bitcoin-NG proposal \cite{eyal2016bitcoin} simplifies the leader election process in Bitcoin by dividing time into a sequence of epochs, and keeping the same block proposer for all blocks within the same epoch. This eliminates latency effects within the epoch, and therefore allows for increased \emph{throughput}. The reduction of latency effects also improves \emph{fairness}. However, the smaller number of nodes that participate in node creation harms \emph{decentralization}. Although the ability of  slot leaders to perform double-spend attacks within epochs is potentially increased, Bitcoin-NG includes measures to counter this such as \qt{poison transactions} and \qt{proofs-of-fraud}, so the total impact on attack resistance is not entirely clear.
\item[Sharding:] In Sharding, the requirement that each node in the network maintains the full transaction ledger is relaxed. Systems that successfully implement sharding have potentially much higher \emph{scalability} and \emph{throughput}. However, in most existing sharding proposals it is much easier to attack a single shard than the entire system, leading to reduced \emph{attack resistance}. Despite some important recent work in this area \cite{zilliqa,Ko18}, practical implementation remains an active research fields with a high potential for future improvements.
\end{description}

In conclusion, the PREStO framework allows for a detailed comparison of different protocols and protocol features that goes far deeper than the safety/decentralization/scalability triad of the blockchain trilemma. As we can see from Table~\ref{tab:two}, this allows us to create a \qt{menu} for protocol designers from which they can choose protocol features that complement or offset each other. As the blockchain ecosystem evolves, this table can be both expanded (in terms of protocols) and refined (in terms of subcategories), making the PREStO a truly dynamic tool for the comparison and evaluation of blockchain protocols.

\begin{table*}[!htb]
\centering
\renewcommand{\arraystretch}{1.1}
\begin{tabular}{lll}
\rowdark
\multicolumn{2}{l}{\sc\b PRESTO Framework} & \multicolumn{1}{c}{\sc \b Design of Blockchain Consensus Layer}\\
\rowdark
&& \multicolumn{1}{c}{Research Challenges -- Opportunities} \\\\[-0.2cm]
\rowlight\sc \b Optimality&
\begin{minipage}[t]{0.4\textwidth}
\begin{itemize}[label=\tabitem]
\item Liveness
\item Safety
\item Scope
\item Privacy features: public/private, permissioned/-less
\end{itemize}
\end{minipage}
&
\begin{minipage}[t]{0.4\textwidth}
\begin{itemize}[label=\tabitem]
\item Selection of design/architecture \& Sybil protection (PoW, PoS etc.).
\item Exploring the trade-off between safety and liveness.
\item Secure execution of smart contracts.
\end{itemize}
\end{minipage}
\\\\[-0.2cm]

\sc \b Stability&
\begin{minipage}[t]{0.4\textwidth}
\begin{itemize}[label=\tabitem]
\item Incentive compatibility: \\Participation, Operations, Applications
\item Decentralization: Entry barriers, Distribution of resources
\item Fairness: reward allocation, voting-decision making 
\end{itemize}
\end{minipage}
&
\begin{minipage}[t]{0.4\textwidth}
\begin{itemize}[label=\tabitem]
\item Design of incentive compatible mechanisms.
\item Protection against adversarial behavior.
\item Motivate decentralization, fair distribution of resources.
\end{itemize}
\end{minipage}
\\\\[-0.2cm]

\rowlight\sc \b Efficiency &
\begin{minipage}[t]{0.4\textwidth}
\begin{itemize}[label=\tabitem]
\item Scalability: positive scale effects
\item Throughput rate
\item Economy of resources/ energy consumption
\item Benchmarking to centralized solutions
\end{itemize}
\end{minipage}
&
\begin{minipage}[t]{0.4\textwidth}
\begin{itemize}[label=\tabitem]
\item Design of scalable properties. 
\item Reduction of energy footprint.
\item Compare blockchain to conventional solutions.
\end{itemize}
\end{minipage}
\\\\[-0.2cm]

\sc \b Robustness &
\begin{minipage}[t]{0.4\textwidth}
\begin{itemize}[label=\tabitem]
\item Tolerance of perturbed assumptions/irrational behaviour
\item Out-of-Protocol Incentives
\item Resilience to attacks:\\ Incentives -- Network -- Cryptographic level
\end{itemize}
\end{minipage}
&
\begin{minipage}[t]{0.4\textwidth}
\begin{itemize}[label=\tabitem]
\item Protection against collusions, Goldfinger attacks.
\item Equilibration in elaborate adversarial models.
\end{itemize}
\end{minipage}
\\\\[-0.2cm]

\rowlight\sc \b Persistence &
\begin{minipage}[t]{0.4\textwidth}
\begin{itemize}[label=\tabitem]
\item Weak/strong persistent properties 
\item Large scale or majority attacks
\item Recovery mechanisms: rare events
\item Governance \& long-term sustainability 
\end{itemize}
\end{minipage}
&
\begin{minipage}[t]{0.4\textwidth}
\begin{itemize}[label=\tabitem]
\item Defense against 51\% attacks, large network partitions
\item Blockchain-Trilemma
\item Design of sustainable blockchains 
\item Decision of governance schemes
\end{itemize}
\end{minipage}\\\\[-0.2cm]
\bottomline
\end{tabular}\vspace*{0.2cm}
\caption{Challenges and current research in the design of Blockchain Protocols based on the PRESTO Framework.}\vspace{-0.3cm}
\label{tab:challenges}
\end{table*}

\subsection{Identifying Research Challenges \& Opportunities}
\label{sec:challenges}
To provide some insights into possibilities for future work, we summarize the categories and subcategories of PREStO and use them to identify research challenges and opportunities in \Cref{tab:challenges}. We elaborate on this in the following.
\begin{description}[noitemsep,leftmargin=*]
\item[Optimality:] In its current stage, the blockchain ecosystem strives to transition to alternative consensus mechanisms that will retain the success of Nakamoto consensus  (which includes PoW) while reducing its energy footprint, \cite{Haz19}. Providing formal guarantees of safety and liveness and testing these new consensus mechanisms in large-scale practical settings is an ongoing challenge. The enhanced ability of the next generation of blockchains to enable and secure the widespread execution of smart contracts only adds to the complexity of this already difficult puzzle.   
\item[Stability:] A large part of the recent blockchain literature has focused on analyzing the incentives in traditional PoW protocols. However, many questions still remain unanswered. Are (virtual) miners motivated to support the network's safety during its ups and downs? What are the vulnerabilities -- at an incentives level -- of the newly proposed PoS protocols and what are the optimal reward schemes that will safeguard their success? Among others, recent advances in theoretical research highlight that centralization and irregular supply of mining power are two threats that are inherent to PoW protocols, \cite{Arn18,Fia19,Gor19}. Partially, these problems stem from cost asymmetries and economies of scale that manifest in energy-consuming consensus mechanisms. Does PoS remedy these problems and is it indeed the next step in blockchain evolution? 
\item[Efficiency:] One pressing -- if not the single most important -- challenge of the blockchain ecosystem is the issue of scalability. Nakamoto's PoW proved unexpectedly successful, yet an important hurdle in Bitcoin's mass adoption is that it does not scale. In particular, even with high energy consumption and increasing environmental externalities, the Bitcoin network cannot process a satisfactory number of transactions per second to compete with the established means of digital payments, like VISA and internet banking. The design of blockchains with scalable properties -- in terms of communication overhead, transaction throughput and smart contract execution -- remains the cornerstone of research in the Efficiency category.
\item[Robustness:] To address the problems that arise in terms of Robustness, the blockchain ecosystem first needs to establish solutions that satisfy Stability and Efficiency. The challenges for robustness will push these solutions to their limits and shift the research towards incorporating these systems to every day life. This is where the dynamic nature of PREStO shines most. Given the state of the art when Stability and Efficiency will have been achieved -- if ever -- the still open questions and challenges can appear as new categories in the framework's Robustness axis. This perspective supports PREStO's current development as a dynamic tool whose purpose is to aid the communication between the increasingly diverse stakeholders of the blockchain community. 
\item[Persistence:] The open challenges in this category will also become more relevant at more mature stages in the mass adoption of blockchain technology. With the exact nature of next-generation blockchains and their main applications still unclear, the issues of governance and long-term sustainability will remain at the forefront of blockchain research. Are public blockchains indeed going to support global payment systems and currencies? And if blockchains and cryptocurrencies indeed succeed in this task, what will be the stance of governments, legislators, regulators and policy makers? How will the major technology firms react to the new technology and how will the existing banking system adapt to the changing reality?
\end{description}
The above provides a non-exhaustive list of the open problems that are currently puzzling the blockchain community. Yet, it highlights the comprehensive coverage of the proposed PREStO framework. Owning to its modular structure, PREStO can be expanded or modified to accommodate advances or additional research opportunities in the future of blockchain protocols. Accordingly, it can be used to track the evolution of the blockchain ecosystem and structure the communication between its diverse participants who range from protocol designers, technology experts and end users to academics, corporate managers and strategic investors.

\section{Conclusions \& Future Research}
\label{sec:conclusions}
Summing up, the PRESTO framework sees protocols as multi-dimensional objects with the following cascade of goals. First, optimality requires that the protocol solves the problem that it is defined to address, otherwise there is no good reason to deploy it and the designer should go back to the drawing board. Second, stability aims to ensure that self-interested agents have an incentive to follow and implement the protocol, i.e., that the protocol itself is an equilibrium. If not, the agents will deviate from it and the deployed protocol will behave unpredictably in practice. Next, efficiency requires that resources are used as efficiently as possible (e.g. time, space, network bandwidth, energy, randomness, etc.). Given an optimal, stable and efficient protocol, the next steps are to consider more elaborate behavioral models from the perspective of the agents. These entail robustness and persistence which measure the resilience of the established equilibria in less idealized settings, and the performance of the blockchain in highly perturbed conditions, respectively.\par
The exploration of these trade-offs is an area for multidisciplinary research that relies on the synthesis of ideas from game theory, cryptography and theoretical computer science. In this direction, PREStO can be used as a dynamic framework to structure the communication between researchers with diverse backgrounds and to accommodate increasingly more elaborate features of future blockchain protocols.

\bibliographystyle{abbrv}
\bibliography{presto_survey_bib}

\begin{thebibliography}{100}

\bibitem{Ab18}
J.~Abadi and M.~Brunnermeier.
\newblock {Blockchain Economics}.
\newblock Working Paper 25407, National Bureau of Economic Research, December
  2018.

\bibitem{Ab06}
I.~Abraham, D.~Dolev, R.~Gonen, and J.~Halpern.
\newblock {Distributed Computing Meets Game Theory: Robust Mechanisms for
  Rational Secret Sharing and Multiparty Computation}.
\newblock In {\em Proceedings of the Twenty-fifth Annual ACM Symposium on
  Principles of Distributed Computing}, PODC '06, pages 53--62, New York, NY,
  USA, 2006. ACM.

\bibitem{Ai05}
A.~S. Aiyer, L.~Alvisi, A.~Clement, M.~Dahlin, J.-P. Martin, and C.~Porth.
\newblock {BAR Fault Tolerance for Cooperative Services}.
\newblock In {\em Proceedings of the Twentieth ACM Symposium on Operating
  Systems Principles}, SOSP '05, pages 45--58, New York, NY, USA, 2005. ACM.

\bibitem{Ap16}
M.~Apostolaki, A.~Zohar, and L.~Vanbever.
\newblock {Hijacking Bitcoin: Large-scale Network Attacks on Cryptocurrencies}.
\newblock {\em CoRR}, abs/1605.07524, 2016.

\bibitem{Arn18}
N.~Arnosti and S.~M. Weinberg.
\newblock {Bitcoin: A Natural Oligopoly}.
\newblock In A.~Blum, editor, {\em 10th Innovations in Theoretical Computer
  Science Conference (ITCS 2019)}, volume 124 of {\em Leibniz International
  Proceedings in Informatics (LIPIcs)}, pages 5:1--5:1, Dagstuhl, Germany,
  2018. Schloss Dagstuhl--Leibniz-Zentrum fuer Informatik.

\bibitem{Ar19}
N.~Arnosti and S.~M. Weinberg.
\newblock {Bitcoin: {A} Natural Oligopoly}.
\newblock In {\em 10th Innovations in Theoretical Computer Science Conference,
  {ITCS} 2019, January 10-12, 2019, San Diego, California, {USA}}, pages
  5:1--5:1, 2019.

\bibitem{As16}
G.~Asharov, R.~Canetti, and C.~Hazay.
\newblock {Toward a Game Theoretic View of Secure Computation}.
\newblock {\em J. Cryptol.}, 29(4):879--926, Oct. 2016.

\bibitem{As10}
K.~J. Astrom and R.~M. Murray.
\newblock {\em Feedback Systems: An Introduction for Scientists and Engineers}.
\newblock Princeton University Press, Princeton, NJ, USA, 2010.

\bibitem{At16}
S.~Athey, I.~Parashkevov, V.~Sarukkai, and J.~Xia.
\newblock {Bitcoin Pricing, Adoption, and Usage: Theory and Evidence}.
\newblock Research paper no. 16--42, Stanford University, Graduate School of
  Business, 2016.

\bibitem{At17}
N.~Atzei, M.~Bartoletti, and T.~Cimoli.
\newblock {A Survey of Attacks on Ethereum Smart Contracts SoK}.
\newblock In {\em Proceedings of the 6th International Conference on Principles
  of Security and Trust - Volume 10204}, pages 164--186, New York, NY, USA,
  2017. Springer-Verlag New York, Inc.

\bibitem{azouvi2019sok}
S.~Azouvi and A.~Hicks.
\newblock {SoK: Tools for Game Theoretic Models of Security for
  Cryptocurrencies}.
\newblock {\em arXiv preprint arXiv:1905.08595}, 2019.

\bibitem{Az18}
S.~Azouvi, A.~Hicks, and S.~J. Murdoch.
\newblock {Incentives in Security Protocols}.
\newblock In V.~Maty{\'a}{\v{s}}, P.~{\v{S}}venda, F.~Stajano, B.~Christianson,
  and J.~Anderson, editors, {\em Security Protocols XXVI}, pages 132--141,
  Cham, 2018. Springer International Publishing.

\bibitem{Ba12}
M.~Babaioff, S.~Dobzinski, S.~Oren, and A.~Zohar.
\newblock {On Bitcoin and Red Balloons}.
\newblock In {\em Proceedings of the 13th ACM Conference on Electronic
  Commerce}, EC '12, pages 56--73, New York, NY, USA, 2012. ACM.

\bibitem{Bad18}
C.~Badertscher, J.~Garay, U.~Maurer, D.~Tschudi, and V.~Zikas.
\newblock {But Why Does It Work? A Rational Protocol Design Treatment of
  Bitcoin}.
\newblock In J.~B. Nielsen and V.~Rijmen, editors, {\em Advances in Cryptology
  -- EUROCRYPT 2018}, pages 34--65. Springer International Publishing, 2018.

\bibitem{Ba11}
M.~Balcan, S.~Krehbiel, G.~Piliouras, and J.~Shin.
\newblock Minimally invasive mechanism design: Distributed covering with
  carefully chosen advice.
\newblock {\em To appear in the Proceedings of the 51st IEEE Conference on
  Decision and Control (CDC)}, 2012.

\bibitem{Bal17}
M.~Ball, A.~Rosen, M.~Sabin, and P.~N. Vasudevan.
\newblock {Proofs of Useful Work}.
\newblock Cryptology ePrint Archive, Report 2017/203, 2017.
\newblock [\href{https://eprint.iacr.org/2017/203}{online}].

\bibitem{Ba99}
S.~A. Banducci and J.~A. Karp.
\newblock {Perceptions of Fairness and Support for Proportional
  Representation}.
\newblock {\em Political Behavior}, 21(3):217--238, 1999.

\bibitem{Ba17}
S.~{Bano}, A.~{Sonnino}, M.~{Al-Bassam}, S.~{Azouvi}, P.~{McCorry},
  S.~{Meiklejohn}, and G.~{Danezis}.
\newblock {Consensus in the Age of Blockchains}.
\newblock {\em arXiv e-prints}, page arXiv:1711.03936, 2017.

\bibitem{Ba17a}
M.~Bartoletti, S.~Lande, and A.~S. Podda.
\newblock {A Proof-of-Stake Protocol for Consensus on Bitcoin Subchains}.
\newblock In M.~Brenner, K.~Rohloff, J.~Bonneau, A.~Miller, P.~Y. Ryan,
  V.~Teague, A.~Bracciali, M.~Sala, F.~Pintore, and M.~Jakobsson, editors, {\em
  Financial Cryptography and Data Security}, pages 568--584, Cham, 2017.
  Springer International Publishing.

\bibitem{Ba52}
W.~Baumol.
\newblock {\em Welfare Economics and the Theory of the State}.
\newblock Harvard University Press, Cambridge, Massachusetts, 1952.

\bibitem{Be01}
A.~Ben-Tal and A.~Nemirovski.
\newblock {\em Lectures on modern convex optimization: analysis, algorithms,
  and engineering applications}.
\newblock SIAM, 2001.

\bibitem{Be16}
I.~Bentov, A.~Gabizon, and A.~Mizrahi.
\newblock {Cryptocurrencies Without Proof of Work}.
\newblock In J.~Clark, S.~Meiklejohn, P.~Y. Ryan, D.~Wallach, M.~Brenner, and
  K.~Rohloff, editors, {\em Financial Cryptography and Data Security}, pages
  142--157, Berlin, Heidelberg, 2016. Springer Berlin Heidelberg.

\bibitem{bentov2017tortoise}
I.~Bentov, P.~Hub{\'a}cek, T.~Moran, and A.~Nadler.
\newblock {Tortoise and Hares Consensus: the Meshcash Framework for
  Incentive-Compatible, Scalable Cryptocurrencies.}
\newblock {\em IACR Cryptology ePrint Archive}, 2017:300, 2017.

\bibitem{bentov2016snow}
I.~Bentov, R.~Pass, and E.~Shi.
\newblock {Snow White: Provably Secure Proofs of Stake.}
\newblock {\em IACR Cryptology ePrint Archive}, 2016:919, 2016.

\bibitem{Bi17}
B.~Biais, C.~Bisière, M.~Bouvard, and C.~Casamatta.
\newblock {The blockchain folk theorem}.
\newblock IDEI Working Papers 873, Institut d'Économie Industrielle (IDEI),
  Toulouse, May 2017.

\bibitem{biere2003bounded}
A.~Biere, A.~Cimatti, E.~M. Clarke, O.~Strichman, Y.~Zhu, et~al.
\newblock Bounded model checking.
\newblock {\em Advances in computers}, 58(11):117--148, 2003.

\bibitem{Bo18}
J.~Bonneau.
\newblock {Hostile blockchain takeovers (short paper)}.
\newblock In {\em Proceedings of the 5th IFCA Workshop on Bitcoin and
  Blockchain Research}, 2018.

\bibitem{Bo16}
J.~Bonneau, E.~Felten, S.~Goldfeder, J.~Kroll, and A.~Narayanan.
\newblock {Why Buy When You Can Rent? Bribery Attacks on Bitcoin-Style
  Consensus}.
\newblock In {\em Financial Cryptography Workshops}, 2016.

\bibitem{Bo15}
J.~Bonneau, A.~Miller, J.~Clark, A.~Narayanan, J.~A. Kroll, and E.~W. Felten.
\newblock {SoK: Research Perspectives and Challenges for Bitcoin and
  Cryptocurrencies}.
\newblock In {\em 2015 IEEE Symposium on Security and Privacy}, pages 104--121,
  2015.

\bibitem{Br18}
J.~{Brown-Cohen}, A.~{Narayanan}, C.-A. {Psomas}, and S.~M. {Weinberg}.
\newblock {Formal Barriers to Longest-Chain Proof-of-Stake Protocols}.
\newblock {\em ArXiv e-prints}, 2018.

\bibitem{Brj18}
L.~{Br{\"u}njes}, A.~{Kiayias}, E.~{Koutsoupias}, and A.-P. {Stouka}.
\newblock {Reward Sharing Schemes for Stake Pools}.
\newblock {\em arXiv e-prints}, page arXiv:1807.11218, 2018.

\bibitem{algorandfunds}
{Business Wire}.
\newblock Algorand secures \$62m in funding and announces appointment of
  executive team, 2018.
\newblock Available
  [\href{https://www.businesswire.com/news/home/20181024005053/en/Algorand-Secures-62M-Funding-Announces-Appointment-Executive}{online}].
  [Accessed: 26-2-2019].

\bibitem{Bu17online}
V.~Buterin.
\newblock The triangle of harm, 2017.
\newblock Available
  [\href{https://vitalik.ca/general/2017/07/16/triangle_of_harm.html}{online}].
  [Accessed: 3-9-2018].

\bibitem{Bu18online8}
V.~Buterin.
\newblock {Discouragement Attacks}, 2018.
\newblock Available
  [\href{https://github.com/ethereum/research/blob/master/papers/discouragement/discouragement.pdf}{online}].
  [Accessed: 13-6-2019].

\bibitem{Bu18online7}
V.~Buterin.
\newblock {Ethereum 2.0 spec -- Casper and sharding}, 2018.
\newblock Available
  [\href{https://github.com/ethereum/eth2.0-specs/blob/master/specs/beacon-chain.md}{online}].
  [Accessed: 30-10-2018].

\bibitem{Bu14online}
V.~Buterin et~al.
\newblock A next-generation smart contract and decentralized application
  platform, 2014.
\newblock Available
  [\href{https://whitepaperdatabase.com/ethereum-eth-whitepaper/}{online}].
  [Accessed: 13-4-2018].

\bibitem{buterin2017casper}
V.~Buterin and V.~Griffith.
\newblock Casper the friendly finality gadget.
\newblock {\em arXiv preprint arXiv:1710.09437}, 2017.

\bibitem{Bu19}
V.~Buterin, D.~Reijsbergen, S.~Leonardos, and G.~Piliouras.
\newblock {Incentives in Ethereum's Hybrid Casper Protocol}.
\newblock {\em International Journal of Network Management}, 30(5):e2098, 2020.
\newblock Special Issue of IEEE International Conference on Blockchain and
  Cryptocurrency.

\bibitem{Cac16}
C.~Cachin.
\newblock Architecture of the {H}yperledger blockchain fabric.
\newblock In {\em Workshop on Distributed Cryptocurrencies and Consensus
  Ledgers}, 2016.

\bibitem{Ca17}
C.~Cachin and M.~Vukoli{\'c}.
\newblock {Blockchain Consensus Protocols in the Wild}.
\newblock {\em CoRR}, abs/1707.01873, 2017.

\bibitem{cardano}
{Cardano}.
\newblock [\href{https://www.cardano.org/en/home/}{website}]. [Accessed:
  26-2-2019].

\bibitem{Ca16}
M.~Carlsten, H.~Kalodner, S.~M. Weinberg, and A.~Narayanan.
\newblock {On the Instability of Bitcoin Without the Block Reward}.
\newblock In {\em Proceedings of the 2016 ACM SIGSAC Conference on Computer and
  Communications Security}, CCS '16, pages 154--167, New York, NY, USA, 2016.
  ACM.

\bibitem{Bit19}
B.~Cash.
\newblock Bitcoin cash: Peer-to-peer electronic cash, 2019.
\newblock Availabel [\href{https://www.bitcoincash.org/}{online}. [Accessed:
  15-12-2019].

\bibitem{Ca19}
F.~Casino, T.~K. Dasaklis, and C.~Patsakis.
\newblock A systematic literature review of blockchain-based applications:
  Current status, classification and open issues.
\newblock {\em Telematics and Informatics}, 36:55--81, 2019.

\bibitem{Cat16}
C.~Catalini and J.~S. Gans.
\newblock {Some Simple Economics of the Blockchain}.
\newblock Working Paper 22952, National Bureau of Economic Research, December
  2016.

\bibitem{iinquorum}
F.~Chaparro.
\newblock Banks have a big appetite to join {JPM}organ's blockchain party.
\newblock Business Insider, 2018.
\newblock Available
  [\href{https://www.businessinsider.sg/blockchain-jpmorgan-says-banks-have-big-appetite-to-join-party-2018-2/?r=UK}{online}].
  [Accessed: 25-2-2019].

\bibitem{Chen19}
X.~Chen, C.~Papadimitriou, and T.~Roughgarden.
\newblock {An Axiomatic Approach to Block Rewards}.
\newblock In {\em Proceedings of the 1st ACM Conference on Advances in
  Financial Technologies}, AFT '19, pages 124--131, New York, NY, USA, 2019.
  ACM.

\bibitem{Ch18}
A.~Chepurnoy, V.~Kharin, and D.~Meshkov.
\newblock {A Systematic Approach To Cryptocurrency Fees}.
\newblock Cryptology ePrint Archive, Report 2018/078, 2018.
\newblock [\href{https://eprint.iacr.org/2018/078}{online}].

\bibitem{chia2018rethinking}
V.~Chia, P.~Hartel, Q.~Hum, S.~Ma, G.~Piliouras, D.~Reijsbergen, M.~van
  Staalduinen, and P.~Szalachowski.
\newblock Rethinking blockchain security: Position paper.
\newblock In {\em 1st IEEE International Conference on Blockchain}, 2018.

\bibitem{Chi17}
J.~Chiu and T.~Koeppl.
\newblock {The Economics of Cryptocurrencies -- Bitcoin and Beyond}.
\newblock Working Paper Series 6688, Victoria University of Wellington, School
  of Economics and Finance, 2017.

\bibitem{Ch05}
G.~Christodoulou and E.~Koutsoupias.
\newblock {The Price of Anarchy of Finite Congestion Games}.
\newblock {\em STOC}, pages 67--73, 2005.

\bibitem{coinmarketcap}
{Coin MarketCap}.
\newblock Available [\href{https://coinmarketcap.com/}{online}]. [Accessed:
  25-2-2019].

\bibitem{Co17}
L.~Cong and Z.~He.
\newblock {Blockchain Disruption and Smart Contracts}.
\newblock Working Paper No. w24399. Available at SSRN: 873, NBER, March 2018.

\bibitem{Co19}
M.~{Conti}, A.~{Gangwal}, and M.~{Todero}.
\newblock {Blockchain Trilemma Solver Algorand has Dilemma over Undecidable
  Messages}.
\newblock {\em arXiv e-prints}, page arXiv:1901.10019, Jan 2019.

\bibitem{Co18}
M.~Conti, E.~S. Kumar, C.~Lal, and S.~Ruj.
\newblock {A Survey on Security and Privacy Issues of Bitcoin}.
\newblock {\em IEEE Communications Surveys Tutorials}, 20(4):3416--3452, 2018.

\bibitem{Cr16}
K.~Croman, C.~Decker, I.~Eyal, A.~E. Gencer, A.~Juels, A.~Kosba, A.~Miller,
  P.~Saxena, E.~Shi, E.~G{\"u}n~Sirer, D.~Song, and R.~Wattenhofer.
\newblock {On Scaling Decentralized Blockchains (Position Paper)}.
\newblock In J.~Clark, S.~Meiklejohn, P.~Y. Ryan, D.~Wallach, M.~Brenner, and
  K.~Rohloff, editors, {\em Financial Cryptography and Data Security}, pages
  106--125, Berlin, Heidelberg, 2016. Springer Berlin Heidelberg.

\bibitem{daian2016snow}
P.~Daian, R.~Pass, and E.~Shi.
\newblock {Snow White: Robustly Reconfigurable Consensus and Applications to
  Provably Secure Proof of Stake}.
\newblock Cryptology ePrint Archive, Report 2016/919, 2016.
\newblock [\href{https://eprint.iacr.org/2016/919}{online}].

\bibitem{Sou18}
S.~Das, A.~Kolluri, P.~Saxena, and H.~Yu.
\newblock {On the Security of Blockchain Consensus Protocols (Invited Paper)}.
\newblock In V.~Ganapathy, T.~Jaeger, and R.~Shyamasundar, editors, {\em
  Information Systems Security}, pages 465--480, Cham, 2018. Springer
  International Publishing.

\bibitem{Da06}
C.~Daskalakis, P.~W. Goldberg, and C.~H. Papadimitriou.
\newblock {The Complexity of Computing a Nash Equilibrium}.
\newblock In {\em Proceedings of the Thirty-eighth Annual ACM Symposium on
  Theory of Computing}, STOC '06, pages 71--78, New York, NY, USA, 2006. ACM.

\bibitem{david2018ouroboros}
B.~David, P.~Ga{\v{z}}i, A.~Kiayias, and A.~Russell.
\newblock Ouroboros praos: An adaptively-secure, semi-synchronous
  proof-of-stake blockchain.
\newblock In {\em Annual International Conference on the Theory and
  Applications of Cryptographic Techniques}, pages 66--98. Springer, 2018.

\bibitem{decker2013information}
C.~Decker and R.~Wattenhofer.
\newblock Information propagation in the bitcoin network.
\newblock In {\em IEEE P2P 2013 Proceedings}, pages 1--10. IEEE, 2013.

\bibitem{fidelity}
M.~del Castillo.
\newblock Fidelity launches institutional platform for {B}itcoin and
  {E}thereum.
\newblock Forbes, 2018.
\newblock Available
  [\href{https://www.forbes.com/sites/michaeldelcastillo/2018/10/15/fidelity-launches-institutional-platform-for-bitcoin-and-ethereum/#4e5a766d93c4}{online}].
  [Accessed: 25-2-2019].

\bibitem{mergers}
M.~del Castillo.
\newblock Despite crypto depression, {M\&A} deals set new record.
\newblock Forbes, 2019.
\newblock Available
  [\href{https://www.forbes.com/sites/michaeldelcastillo/2019/02/13/despite-crypto-depression-ma-deals-set-new-record/#1d6b65952444}{online}].
  [Accessed: 25-2-2019].

\bibitem{nasdaq}
M.~del Castillo.
\newblock Nasdaq leads \$20 million investment in enterprise blockchain startup
  {S}ymbiont.
\newblock Forbes, 2019.
\newblock Available
  [\href{https://www.forbes.com/sites/michaeldelcastillo/2019/01/23/exclusive-nasdaq-leads-20-million-investment-in-enterprise-blockchain-startup-symbiont/#18083f3346d1}{online}].
  [Accessed: 25-2-2019].

\bibitem{Dh18}
S.~{Dhamal}, T.~{Chahed}, W.~{Ben-Ameur}, E.~{Altman}, A.~{Sunny}, and
  S.~{Poojary}.
\newblock {A Stochastic Game Framework for Analyzing Computational Investment
  Strategies in Distributed Computing with Application to Blockchain Mining}.
\newblock {\em arXiv e-prints}, page arXiv:1809.03143, 2018.

\bibitem{Di18online}
Digiconomist.
\newblock {Bitcoin Energy Consumption Index}.
\newblock Available
  [\href{https://digiconomist.net/bitcoin-energy-consumption}{online}].
  [Accessed: 3-9-2018].

\bibitem{Di17}
N.~Dimitri.
\newblock {Bitcoin Mining as a Contest}.
\newblock {\em Ledger}, 2(0):31--37, 2017.

\bibitem{Di18}
T.~T.~A. Dinh, R.~Liu, M.~Zhang, G.~Chen, B.~C. Ooi, and J.~Wang.
\newblock {Untangling Blockchain: A Data Processing View of Blockchain
  Systems}.
\newblock {\em IEEE Transactions on Knowledge and Data Engineering},
  30(7):1366--1385, 2018.

\bibitem{Din17}
T.~T.~A. Dinh, J.~Wang, G.~Chen, R.~Liu, B.~C. Ooi, and K.-L. Tan.
\newblock {BLOCKBENCH: A Framework for Analyzing Private Blockchains}.
\newblock In {\em Proceedings of the 2017 ACM International Conference on
  Management of Data}, SIGMOD '17, pages 1085--1100, New York, NY, USA, 2017.
  ACM.

\bibitem{Eos19}
EOS.IO.
\newblock Eosio strategic vision, 2019.
\newblock Available [\href{https://eos.io/strategic-vision/}{online}.
  [Accessed: 15-12-2019].

\bibitem{Et20online}
{Ethereum}.
\newblock {Sharding FAQ}.
\newblock Available
  [\href{https://github.com/ethereum/wiki/wiki/Sharding-FAQ}{online}].
  [Accessed: 13-2-2020].

\bibitem{Et18online2}
{Ethereum}.
\newblock {Sharding Roadmap}.
\newblock Available
  [\href{https://github.com/ethereum/wiki/wiki/Sharding-roadmap}{online}].
  [Accessed: 14-9-2018].

\bibitem{Ey15}
I.~Eyal.
\newblock {The Miner's Dilemma}.
\newblock In {\em Proceedings of the 2015 IEEE Symposium on Security and
  Privacy}, SP '15, pages 89--103, Washington, DC, USA, 2015. IEEE Computer
  Society.

\bibitem{eyal2016bitcoin}
I.~Eyal, A.~E. Gencer, E.~G. Sirer, and R.~Van~Renesse.
\newblock Bitcoin-ng: A scalable blockchain protocol.
\newblock In {\em 13th {USENIX} Symposium on Networked Systems Design and
  Implementation ({NSDI})}, pages 45--59, 2016.

\bibitem{Ey14}
I.~Eyal and E.~Sirer.
\newblock {Majority Is Not Enough: Bitcoin Mining Is Vulnerable}.
\newblock In N.~Christin and R.~Safavi-Naini, editors, {\em Financial
  Cryptography and Data Security}, pages 436--454, Berlin, Heidelberg, 2014.
  Springer Berlin Heidelberg.

\bibitem{Fa18}
G.~{Fanti}, L.~{Kogan}, S.~{Oh}, K.~{Ruan}, P.~{Viswanath}, and G.~{Wang}.
\newblock {Compounding of Wealth in Proof-of-Stake Cryptocurrencies}.
\newblock {\em ArXiv e-prints}, 2018.

\bibitem{ghash}
C.~Farivar.
\newblock {Bitcoin pool GHash.io commits to 40\% hashrate limit after its 51\%
  breach}.
\newblock
  [\href{https://arstechnica.com/information-technology/2014/07/bitcoin-pool-ghash-io-commits-to-40-hashrate-limit-after-its-51-breach/}{online}],
  2014.

\bibitem{Fe98}
D.~Felsenthal and M.~Machover.
\newblock {\em {The Measurement of Voting Power: Theory and Practice, Problems
  and Paradoxes}}.
\newblock Edward Elgar Publishing, Inc., 1998.

\bibitem{Fe18}
S.~{Feng}, W.~{Wang}, Z.~{Xiong}, D.~{Niyato}, P.~{Wang}, and S.~{Shuxun Wang}.
\newblock {On Cyber Risk Management of Blockchain Networks: A Game Theoretic
  Approach}.
\newblock {\em arXiv e-prints}, page arXiv:1804.10412, Apr. 2018.

\bibitem{ferrag2018blockchain}
M.~A. Ferrag, M.~Derdour, M.~Mukherjee, A.~Derhab, L.~Maglaras, and H.~Janicke.
\newblock Blockchain technologies for the internet of things: Research issues
  and challenges.
\newblock {\em IEEE Internet of Things Journal}, 2018.

\bibitem{Fia19}
A.~Fiat, A.~Karlin, E.~Koutsoupias, and C.~Papadimitriou.
\newblock {Energy Equilibria in Proof-of-Work Mining}.
\newblock In {\em Proceedings of the 2019 ACM Conference on Economics and
  Computation}, EC '19, pages 489--502, New York, NY, USA, 2019. ACM.

\bibitem{Fi17}
B.~Fisch, R.~Pass, and A.~Shelat.
\newblock {Socially Optimal Mining Pools}.
\newblock In N.~R.~Devanur and P.~Lu, editors, {\em Web and Internet
  Economics}, pages 205--218, Cham, 2017. Springer International Publishing.

\bibitem{Ga13}
J.~Garay, J.~Katz, U.~Maurer, B.~Tackmann, and V.~Zikas.
\newblock {Rational Protocol Design: Cryptography against Incentive-Driven
  Adversaries}.
\newblock In {\em 2013 IEEE 54th Annual Symposium on Foundations of Computer
  Science}, pages 648--657, Oct 2013.

\bibitem{Ga15}
J.~Garay, A.~Kiayias, and N.~Leonardos.
\newblock {The Bitcoin Backbone Protocol: Analysis and Applications}.
\newblock In {\em Annual International Conference on the Theory and
  Applications of Cryptographic Techniques}, pages 281--310. Springer, 2015.

\bibitem{Ga17}
J.~Garay, A.~Kiayias, and N.~Leonardos.
\newblock {The Bitcoin Backbone Protocol with Chains of Variable Difficulty}.
\newblock In J.~Katz and H.~Shacham, editors, {\em Advances in Cryptology --
  CRYPTO 2017}, pages 291--323, Cham, 2017. Springer International Publishing.

\bibitem{Gar18}
J.~A. Garay and A.~Kiayias.
\newblock {SoK: A Consensus Taxonomy in the Blockchain Era}.
\newblock {\em IACR Cryptology ePrint Archive}, 2018:754, 2018.

\bibitem{Leo18}
J.~A. Garay, A.~Kiayias, N.~Leonardos, and G.~Panagiotakos.
\newblock {Bootstrapping the Blockchain, with Applications to Consensus and
  Fast PKI Setup}.
\newblock In {\em Public Key Cryptography}, 2018.
\newblock [\href{https://eprint.iacr.org/2016/991}{online}].

\bibitem{ibm}
A.~Garcia.
\newblock {{IBM} is betting big on blockchain technology. Is it worth the
  risk?}
\newblock CNN Business, 2018.
\newblock Available
  [\href{https://money.cnn.com/2018/09/06/technology/ibm-blockchain-gamble/index.html}{online}].
  [Accessed: 25-2-2019].

\bibitem{Ga11}
V.~K. Garg and J.~Bridgman.
\newblock {The Weighted Byzantine Agreement Problem}.
\newblock In {\em 2011 IEEE International Parallel Distributed Processing
  Symposium}, pages 524--531, May 2011.

\bibitem{Ga18}
P.~Gazi, A.~Kiayias, and A.~Russell.
\newblock {Stake-Bleeding Attacks on Proof-of-Stake Blockchains}.
\newblock {\em {2018 Crypto Valley Conference on Blockchain Technology
  (CVCBT)}}, pages 85--92, 2018.

\bibitem{Gaz18}
P.~Gazi, A.~Kiayias, and D.~Zindros.
\newblock {Proof-of-Stake Sidechains}.
\newblock Cryptology ePrint Archive, Report 2018/1239, 2018.
\newblock [\href{https://eprint.iacr.org/2018/1239}{online}].

\bibitem{Ge14}
A.~Gervais, G.~O. Karame, V.~Capkun, and S.~Capkun.
\newblock {Is Bitcoin a Decentralized Currency?}
\newblock {\em IEEE Security \& Privacy}, 12(3):54--60, May 2014.

\bibitem{Ge16}
A.~Gervais, G.~O. Karame, K.~W\"{u}st, V.~Glykantzis, H.~Ritzdorf, and
  S.~Capkun.
\newblock {On the Security and Performance of Proof of Work Blockchains}.
\newblock In {\em Proceedings of the 2016 ACM SIGSAC Conference on Computer and
  Communications Security}, CCS '16, pages 3--16, New York, NY, USA, 2016. ACM.

\bibitem{Gi17}
Y.~Gilad, R.~Hemo, S.~Micali, G.~Vlachos, and N.~Zeldovich.
\newblock Algorand: Scaling byzantine agreements for cryptocurrencies.
\newblock In {\em Proceedings of the 26th Symposium on Operating Systems
  Principles}, pages 51--68. ACM, 2017.

\bibitem{gilbert2002brewer}
S.~Gilbert and N.~Lynch.
\newblock Brewer's conjecture and the feasibility of consistent, available,
  partition-tolerant web services.
\newblock {\em Acm Sigact News}, 33(2):51--59, 2002.

\bibitem{Gj16}
H.~Gjermundr{\o}d, K.~Chalkias, and I.~Dionysiou.
\newblock {Going Beyond the Coinbase Transaction Fee: Alternative Reward
  Schemes for Miners in Blockchain Systems}.
\newblock In {\em Proceedings of the 20th Pan-Hellenic Conference on
  Informatics}, PCI '16, pages 35:1--35:4, New York, NY, USA, 2016. ACM.

\bibitem{goldwasser1989knowledge}
S.~Goldwasser, S.~Micali, and C.~Rackoff.
\newblock The knowledge complexity of interactive proof systems.
\newblock {\em SIAM Journal on computing}, 18(1):186--208, 1989.

\bibitem{Gor19}
G.~Goren and A.~Spiegelman.
\newblock Mind the mining.
\newblock In {\em Proceedings of the 2019 ACM Conference on Economics and
  Computation}, EC '19, pages 475--487, New York, NY, USA, 2019. ACM.

\bibitem{Gr14}
R.~Gradwohl and O.~Reingold.
\newblock Fault tolerance in large games.
\newblock {\em Games and Economic Behavior}, 86:438--457, 2014.

\bibitem{Gu18}
R.~Guerraoui and J.~Wang.
\newblock On the unfairness of blockchain.
\newblock Technical report, EPFL Scientific Publications, 2018.

\bibitem{Ha11}
J.~Y. Halpern.
\newblock {Beyond Nash Equilibrium: Solution Concepts for the 21st Century}.
\newblock In J.~S. Baras, J.~Katz, and E.~Altman, editors, {\em Decision and
  Game Theory for Security}, pages 1--3, Berlin, Heidelberg, 2011. Springer
  Berlin Heidelberg.

\bibitem{Haz19}
S.~S. Hazari and Q.~H. Mahmoud.
\newblock Comparative evaluation of consensus mechanisms in cryptocurrencies.
\newblock {\em Internet Technology Letters}, 2(3):e100, 2019.

\bibitem{He15}
E.~Heilman, A.~Kendler, A.~Zohar, and S.~Goldberg.
\newblock {Eclipse Attacks on Bitcoin{\textquoteright}s Peer-to-Peer Network}.
\newblock In {\em 24th {USENIX} Security Symposium ({USENIX} Security 15)},
  pages 129--144, Washington, D.C., 2015. {USENIX} Association.

\bibitem{Al18online}
A.~Hertig.
\newblock {Blockchain's Once-Feared 51\% Attack Is Now Becoming Regular}.
\newblock Coindesk.com, 2018.
\newblock Available
  [\href{https://www.coindesk.com/blockchains-feared-51-attack-now-becoming-regular}{online}].
  [Accessed: 26-02-2019].

\bibitem{Ho98}
J.~Hofbauer and K.~Sigmund.
\newblock {\em Evolutionary Games and Population Dynamics}.
\newblock Cambridge University Press, Cambridge, 1998.

\bibitem{Ho19online}
C.~Hoskinson.
\newblock {Ethereum Cofounder Says Blockchain Presents \qt{Governance Crisis}},
  2019.
\newblock Available
  [\href{http://fortune.com/2019/04/08/ethereum-cofounder-governance-charles-hoskinson/}{online}].
  [Accessed: 11-4-2019].

\bibitem{Hu18}
Z.~Hu and J.~Zhang.
\newblock {Toward General Robustness Evaluation of Incentive Mechanism Against
  Bounded Rationality}.
\newblock {\em IEEE Transactions on Computational Social Systems},
  5(3):698--712, Sep. 2018.

\bibitem{hlwalmart}
{HyperLedger}.
\newblock Walmart turns to blockchain (and {H}yperledger) to take on food
  traceability and safety, 2018.
\newblock Available
  [\href{https://www.hyperledger.org/blog/2019/02/21/walmart-turns-to-blockchain-and-hyperledger-to-take-on-food-traceability-and-safety}{online}].
  [Accessed: 25-2-2019].

\bibitem{Hyp19}
{HyperLedger}.
\newblock Hyyperledger, 2019.
\newblock Available [\href{https://www.hyperledger.org}{online}]. [Accessed:
  15-12-2019].

\bibitem{Im10}
N.~Immorlica, E.~Markakis, and G.~Piliouras.
\newblock {Coalition Formation and Price of Anarchy in Cournot Oligopolies}.
\newblock In A.~Saberi, editor, {\em Workshop on Internet and Network Economics
  (WINE)}, pages 270--281, Berlin, Heidelberg, 2010. Springer Berlin
  Heidelberg.

\bibitem{istanbulbft}
{Istanbul Byzantine Fault Tolerance}.
\newblock \href{https://github.com/ethereum/EIPs/issues/650}{website}.
  [Accessed: 13-6-2019].

\bibitem{Ja17}
A.~D. Jaggard, N.~Lutz, M.~Schapira, and R.~N. Wright.
\newblock {Dynamics at the Boundary of Game Theory and Distributed Computing}.
\newblock {\em ACM Trans. Econ. Comput.}, 5(3):15:1--15:20, 2017.

\bibitem{Je19online}
G.~Jenkinson.
\newblock {Ethereum Classic 51\% Attack -- The Reality of Proof-of-Work}.
\newblock Cointelegraph.com, 2019.
\newblock Available
  [\href{https://cointelegraph.com/news/ethereum-classic-51-attack-the-reality-of-proof-of-work}{online}].
  [Accessed: 26-02-2019].

\bibitem{Jia2018}
Y.~Jia.
\newblock Op ed: The many faces of sharding for blockchain scalability.
\newblock Bitcoin Magazine, 2018.
\newblock
  [\href{https://bitcoinmagazine.com/articles/op-ed-many-faces-sharding-blockchain-scalability}{online}].

\bibitem{Jo14}
B.~Johnson, A.~Laszka, J.~Grossklags, M.~Vasek, and T.~Moore.
\newblock {Game-Theoretic Analysis of DDoS Attacks Against Bitcoin Mining
  Pools}.
\newblock In R.~B{\"o}hme, M.~Brenner, T.~Moore, and M.~Smith, editors, {\em
  Financial Cryptography and Data Security}, pages 72--86, Berlin, Heidelberg,
  2014. Springer Berlin Heidelberg.

\bibitem{kappos2018empirical}
G.~Kappos, H.~Yousaf, M.~Maller, and S.~Meiklejohn.
\newblock An empirical analysis of anonymity in {Zcash}.
\newblock In {\em 27th {USENIX} Security Symposium}, pages 463--477, 2018.

\bibitem{Ka17}
A.~Kaushik, A.~Choudhary, C.~Ektare, D.~Thomas, and S.~Akram.
\newblock {Blockchain -- Literature survey}.
\newblock In {\em 2017 2nd IEEE International Conference on Recent Trends in
  Electronics, Information Communication Technology (RTEICT)}, pages
  2145--2148, 2017.

\bibitem{Ki16}
A.~Kiayias, E.~Koutsoupias, M.~Kyropoulou, and Y.~Tselekounis.
\newblock {Blockchain Mining Games}.
\newblock In {\em Proceedings of the 2016 ACM Conference on Economics and
  Computation}, pages 365--382. ACM, 2016.

\bibitem{kiayias2017ouroboros}
A.~Kiayias, A.~Russell, B.~David, and R.~Oliynykov.
\newblock Ouroboros: A provably secure proof-of-stake blockchain protocol.
\newblock In {\em Annual International Cryptology Conference}, pages 357--388.
  Springer, 2017.

\bibitem{Ki12}
S.~King and S.~Nadal.
\newblock {PPCoin: Peer-to-Peer Crypto-Currency with Proof-of-Stake}, 2012.

\bibitem{Kok20}
C.~Koki, S.~Leonardos, and G.~Piliouras.
\newblock {Do Cryptocurrency Prices Camouflage Latent Economic Effects? A
  Bayesian Hidden Markov Approach}.
\newblock {\em Future Internet}, 12(3), 2020.
\newblock {Special Issue: Selected Papers from the 3rd Annual Decentralized
  Conference (Best Paper Award)}.

\bibitem{Ko18}
E.~Kokoris-Kogias, P.~Jovanovic, L.~Gasser, N.~Gailly, E.~Syta, and B.~Ford.
\newblock Omniledger: A secure, scale-out, decentralized ledger via sharding.
\newblock In {\em 2018 IEEE Symposium on Security and Privacy (SP)}, pages
  583--598. IEEE, 2018.

\bibitem{Ko99}
E.~Koutsoupias and C.~H. Papadimitriou.
\newblock {Worst-case Equilibria}.
\newblock In {\em Annual Symposium on Theoretical Aspects of Computer Science
  (STACS)}, pages 404--413. Springer-Verlag, 1999.

\bibitem{Kr13}
J.~A. Kroll, I.~C. Davey, and E.~W. Felten.
\newblock {The Economics of Bitcoin Mining, or Bitcoin in the Presence of
  Adversaries}.
\newblock In {\em Workshop on the Economics of Information Security}, volume
  2013, page~11, 2013.

\bibitem{kshetri20181}
N.~Kshetri.
\newblock Blockchain's roles in meeting key supply chain management objectives.
\newblock {\em International Journal of Information Management}, 39:80--89,
  2018.

\bibitem{Kw14}
J.~Kwon.
\newblock Tendermint: Consensus without mining, 2014.

\bibitem{lamport1998part}
L.~Lamport et~al.
\newblock The part-time parliament.
\newblock {\em ACM Transactions on Computer Systems}, 1998.

\bibitem{lamport1982byzantine}
L.~Lamport, R.~Shostak, and M.~Pease.
\newblock The byzantine generals problem.
\newblock {\em ACM Transactions on Programming Languages and Systems (TOPLAS)},
  4(3):382--401, 1982.

\bibitem{La15}
A.~Laszka, B.~Johnson, and J.~Grossklags.
\newblock {When Bitcoin Mining Pools Run Dry}.
\newblock In M.~Brenner, N.~Christin, B.~Johnson, and K.~Rohloff, editors, {\em
  Financial Cryptography and Data Security}, pages 63--77, Berlin, Heidelberg,
  2015. Springer Berlin Heidelberg.

\bibitem{Le11}
C.~Lees.
\newblock {Coalition Formation and the German Party System}.
\newblock {\em German Politics}, 20(1):146--163, 2011.

\bibitem{Leo19}
N.~Leonardos, S.~Leonardos, and G.~Piliouras.
\newblock {Oceanic Games: Centralization Risks and Incentives in Blockchain
  Mining}.
\newblock In P.~Pardalos, I.~Kotsireas, Y.~Guo, and W.~Knottenbelt, editors,
  {\em Mathematical Research for Blockchain Economy}, pages 183--199, Cham,
  2020. Springer International Publishing.
\newblock Best Paper Award.

\bibitem{Le19}
S.~Leonardos, D.~Reijsbergen, and G.~Piliouras.
\newblock {Weighted Voting on the Blockchain: Improving Consensus in Proof of
  Stake Protocols}.
\newblock {\em International Journal of Network Management}, 30(5):e2093, 2020.
\newblock Special Issue of IEEE International Conference on Blockchain and
  Cryptocurrency (Best Paper Award).

\bibitem{Hu06}
S.~R. Leonid~Hurwicz.
\newblock {\em {Designing Economic Mechanisms}}.
\newblock Cambridge University Press, New York, USA, 2006.

\bibitem{Le15}
Y.~Lewenberg, Y.~Bachrach, Y.~Sompolinsky, A.~Zohar, and J.~S. Rosenschein.
\newblock {Bitcoin Mining Pools: A Cooperative Game Theoretic Analysis}.
\newblock In {\em Proceedings of the 2015 International Conference on
  Autonomous Agents and Multiagent Systems}, AAMAS '15, pages 919--927,
  Richland, SC, 2015. International Foundation for Autonomous Agents and
  Multiagent Systems.

\bibitem{Lit19}
Litecoin.
\newblock Litecoin, 2019.
\newblock Availabel [\href{https://litecoin.com/en/}{online}. [Accessed:
  15-12-2019].

\bibitem{Li18}
X.~Liu, W.~Wang, D.~Niyato, N.~Zhao, and P.~Wang.
\newblock {Evolutionary Game for Mining Pool Selection in Blockchain Networks}.
\newblock {\em {IEEE Wireless Communications Letters}}, 7(5):760--763, Oct
  2018.

\bibitem{Li16}
Y.~Liu, J.~Zhang, B.~An, and S.~Sen.
\newblock {A simulation framework for measuring robustness of incentive
  mechanisms and its implementation in reputation systems}.
\newblock {\em Autonomous Agents and Multi-Agent Systems}, 30(4):581--600, Jul
  2016.

\bibitem{Del19online}
D.~LLP.
\newblock {Blockchain Enigma. Paradox. Opportunity}, 2019.
\newblock Available
  [\href{https://www2.deloitte.com/ch/en/pages/innovation/articles/blockchain.html}{online}].
  [Accessed: 10-4-2019].

\bibitem{Luu16}
L.~Luu, V.~Narayanan, C.~Zheng, K.~Baweja, S.~Gilbert, and P.~Saxena.
\newblock A secure sharding protocol for open blockchains.
\newblock In {\em Proceedings of the 2016 ACM SIGSAC Conference on Computer and
  Communications Security}, pages 17--30. ACM, 2016.

\bibitem{Lu15b}
L.~Luu, R.~Saha, I.~Parameshwaran, P.~Saxena, and A.~Hobor.
\newblock {On Power Splitting Games in Distributed Computation: The Case of
  Bitcoin Pooled Mining}.
\newblock In {\em 2015 IEEE 28th Computer Security Foundations Symposium},
  pages 397--411, 2015.

\bibitem{Lu15a}
L.~Luu, J.~Teutsch, R.~Kulkarni, and P.~Saxena.
\newblock {Demystifying Incentives in the Consensus Computer}.
\newblock In {\em Proceedings of the 22Nd ACM SIGSAC Conference on Computer and
  Communications Security}, CCS '15, pages 706--719, New York, NY, USA, 2015.
  ACM.

\bibitem{Lu17}
L.~Luu, Y.~Velner, J.~Teutsch, and P.~Saxena.
\newblock {SMARTPOOL: Practical Decentralized Pooled Mining}.
\newblock In {\em Proceedings of the 26th USENIX Conference on Security
  Symposium}, SEC'17, pages 1409--1426, Berkeley, CA, USA, 2017. USENIX
  Association.

\bibitem{mccorry2018smart}
P.~McCorry, A.~Hicks, and S.~Meiklejohn.
\newblock Smart contracts for bribing miners.
\newblock In {\em International Conference on Financial Cryptography and Data
  Security}, pages 3--18. Springer, 2018.

\bibitem{merkle1987digital}
R.~C. Merkle.
\newblock A digital signature based on a conventional encryption function.
\newblock In {\em Conference on the theory and application of cryptographic
  techniques}, pages 369--378. Springer, 1987.

\bibitem{mettler2016blockchain}
M.~Mettler.
\newblock Blockchain technology in healthcare: The revolution starts here.
\newblock In {\em 2016 IEEE 18th International Conference on e-Health
  Networking, Applications and Services (Healthcom)}, pages 1--3. IEEE, 2016.

\bibitem{Mi13online}
A.~Miller.
\newblock Feather-forks: enforcing a blacklist with sub-50\% hash power, 2013.
\newblock Available
  [\href{https://bitcointalk.org/index.php?topic=312668.0}{online}]. [Accessed:
  3-9-2018].

\bibitem{Mo09}
T.~Moscibroda, S.~Schmid, and R.~Wattenhofer.
\newblock {The Price of Malice: A Game-Theoretic Framework for Malicious
  Behavior in Distributed Systems}.
\newblock {\em Internet Mathematics}, 6(2):125--155, 2009.

\bibitem{My07}
R.~B. Myerson.
\newblock {\em Game Theory}.
\newblock Harvard University Press, Cambridge, Massachusetts, 2007.

\bibitem{Na08}
S.~Nakamoto.
\newblock {Bitcoin: A Peer-to-Peer Electronic Cash System}, 2008.
\newblock Available [\href{https://bitcoin.org/bitcoin.pdf}{online}].
  [Accessed: 14-11-2018].

\bibitem{Na50}
J.~Nash.
\newblock Equilibrium points in n-person games.
\newblock {\em Proceedings of the National Academy of Sciences}, pages 48--49,
  1950.

\bibitem{Nay16}
K.~Nayak, S.~Kumar, A.~Miller, and E.~Shi.
\newblock {Stubborn Mining: Generalizing Selfish Mining and Combining with an
  Eclipse Attack}.
\newblock In {\em 2016 IEEE European Symposium on Security and Privacy (EuroS
  P)}, pages 305--320, March 2016.

\bibitem{Ni07}
N.~Nisan, T.~Roughgarden, E.~Tardos, and V.~V. Vazirani.
\newblock {\em Algorithmic Game Theory}.
\newblock Cambridge University Press, New York, NY, USA, 2007.

\bibitem{No17}
M.~Nojoumian, A.~Golchubian, and L.~Njilla.
\newblock {Incentivizing Blockchain Miners to Avoid Dishonest Mining Strategies
  By a Reputation-Based Paradigm}.
\newblock In {\em IEEE Science and Information Conference}, pages 1118--1134,
  2017.

\bibitem{jpmiin}
L.~Noonan.
\newblock {JPMorgan} widens blockchain payments to more than 75 banks.
\newblock Financial Times, 2018.
\newblock Available
  [\href{https://www.ft.com/content/41bb140e-bc53-11e8-94b2-17176fbf93f5}{online}].
  [Accessed: 25-2-2019].

\bibitem{Od14}
K.~J. O'Dwyer and D.~Malone.
\newblock {Bitcoin mining and its energy footprint}.
\newblock In {\em 25th IET Irish Signals Systems Conference 2014 and 2014
  China-Ireland International Conference on Information and Communications
  Technologies (ISSC 2014/CIICT 2014)}, pages 280--285, June 2014.

\bibitem{ongaro2014search}
D.~Ongaro and J.~Ousterhout.
\newblock In search of an understandable consensus algorithm.
\newblock In {\em 2014 USENIX Annual Technical Conference}, pages 305--319,
  2014.

\bibitem{Pa03}
C.~H. Papadimitriou.
\newblock {\em Computational complexity}.
\newblock John Wiley and Sons Ltd., 2003.

\bibitem{Pa17a}
R.~Pass, L.~Seeman, and A.~Shelat.
\newblock Analysis of the blockchain protocol in asynchronous networks.
\newblock In {\em Annual International Conference on the Theory and
  Applications of Cryptographic Techniques}, pages 643--673. Springer, 2017.

\bibitem{Pa17b}
R.~Pass and E.~Shi.
\newblock {FruitChains: A Fair Blockchain}.
\newblock In {\em Proceedings of the ACM Symposium on Principles of Distributed
  Computing}, PODC '17, pages 315--324, New York, NY, USA, 2017. ACM.

\bibitem{Pi14b}
G.~Piliouras, C.~Nieto-Granda, H.~I. Christensen, and J.~S. Shamma.
\newblock {Persistent Patterns: Multi-agent Learning Beyond Equilibrium and
  Utility}.
\newblock In {\em Proceedings of the 2014 International Conference on
  Autonomous Agents and Multi-agent Systems}, AAMAS '14, pages 181--188,
  Richland, SC, 2014. International Foundation for Autonomous Agents and
  Multiagent Systems.

\bibitem{Pi16}
G.~Piliouras, E.~Nikolova, and J.~S. Shamma.
\newblock {Risk Sensitivity of Price of Anarchy Under Uncertainty}.
\newblock {\em ACM Trans. Econ. Comput.}, 5(1):5:1--5:27, Oct. 2016.

\bibitem{Pi14a}
G.~Piliouras and J.~S. Shamma.
\newblock Optimization despite chaos: Convex relaxations to complex limit sets
  via {P}oincar{\'e} recurrence.
\newblock In {\em Proceedings of the twenty-fifth annual ACM-SIAM Symposium on
  Discrete Algorithms}, pages 861--873. SIAM, 2014.

\bibitem{Pra19}
Prasanna.
\newblock {Blockchain Trilemma: Explained}, 2019.
\newblock Available
  [\href{https://cryptoticker.io/en/blockchain-trilemma-explained/}{online}.
  [Accessed: 15-12-2019].

\bibitem{Pri18online}
Researchers link realism to blockchain’s promise, 2018.
\newblock Available
  [\href{https://www.princeton.edu/news/2018/12/26/researchers-link-realism-blockchains-promise}{online}].
  [Accessed: 10-4-2019].

\bibitem{quorum}
{Quorum}.
\newblock [\href{https://www.jpmorgan.com/global/Quorum}{website}]. [Accessed:
  25-2-2019].

\bibitem{Re14}
L.~Ren.
\newblock {Proof of Stake Velocity: Building the Social Currency of the Digital
  Age}, 2014.

\bibitem{Rh93}
S.~Rhoades.
\newblock {The Herfindahl-Hirschman Index}.
\newblock {\em Federal Reserve Bulletin}, 79:188, 1993.

\bibitem{rivest2001leak}
R.~L. Rivest, A.~Shamir, and Y.~Tauman.
\newblock How to leak a secret.
\newblock In {\em International Conference on the Theory and Application of
  Cryptology and Information Security}, pages 552--565. Springer, 2001.

\bibitem{Ro18online}
J.~J. Roberts.
\newblock {Bitcoin Spinoff Hacked in Rare \qt{51\% Attack}}.
\newblock Fortune, 2018.
\newblock Available
  [\href{http://fortune.com/2018/05/29/bitcoin-gold-hack/}{online}]. [Accessed:
  15-03-2019].

\bibitem{rosenfeld2011analysis}
M.~Rosenfeld.
\newblock Analysis of bitcoin pooled mining reward systems.
\newblock {\em arXiv preprint arXiv:1112.4980}, 2011.

\bibitem{Ro09}
T.~Roughgarden.
\newblock {Intrinsic robustness of the price of anarchy}.
\newblock In {\em Proceedings of the 41st Annual {ACM} Symposium on Theory of
  Computing, {STOC} 2009, Bethesda, MD, USA, May 31 - June 2, 2009}, pages
  513--522, 2009.

\bibitem{Sal17}
F.~A. Saleh.
\newblock {Blockchain Without Waste: Proof-of-Stake}, 2017.

\bibitem{Sa17}
A.~Sapirshtein, Y.~Sompolinsky, and A.~Zohar.
\newblock {Optimal Selfish Mining Strategies in Bitcoin}.
\newblock In J.~Grossklags and B.~Preneel, editors, {\em Financial Cryptography
  and Data Security}, pages 515--532, Berlin, Heidelberg, 2017. Springer Berlin
  Heidelberg.

\bibitem{jpmcoin}
A.~Saxena.
\newblock {JPMorgan Chase} to create digital coins using blockchain for
  payments.
\newblock Reuters, 2019.
\newblock Available
  [\href{https://www.reuters.com/article/us-jp-morgan-blockchain/jpmorgan-chase-to-create-digital-coins-using-blockchain-for-payments-idUSKCN1Q321P}{online}].
  [Accessed: 25-2-2019].

\bibitem{Sc05}
T.~C. Schelling.
\newblock {\em The Strategy of Conflict}.
\newblock Harvard University Press, Cambridge, Massachusetts, 2005.

\bibitem{Sc99}
K.~M. Schmidt and E.~Fehr.
\newblock {A Theory of Fairness, Competition, and Cooperation*}.
\newblock {\em The Quarterly Journal of Economics}, 114(3):817--868, 08 1999.

\bibitem{Sc17}
O.~Schrijvers, J.~Bonneau, D.~Boneh, and T.~Roughgarden.
\newblock {Incentive Compatibility of Bitcoin Mining Pool Reward Functions}.
\newblock In J.~Grossklags and B.~Preneel, editors, {\em Financial Cryptography
  and Data Security}, pages 477--498, Berlin, Heidelberg, 2017. Springer Berlin
  Heidelberg.

\bibitem{Se18}
S.~Seang and D.~Torre.
\newblock {Proof of Work and Proof of Stake Consensus Protocols: A Blockchain
  Application for Local Complementary Currencies}, 2018.

\bibitem{sergey2018scilla}
I.~Sergey, A.~Kumar, and A.~Hobor.
\newblock Scilla: a smart contract intermediate-level language.
\newblock {\em arXiv preprint arXiv:1801.00687}, 2018.

\bibitem{shahaab2019applicability}
A.~Shahaab, B.~Lidgey, C.~Hewage, and I.~Khan.
\newblock Applicability and appropriateness of distributed ledgers consensus
  protocols in public and private sectors: A systematic review.
\newblock {\em IEEE Access}, 7:43622--43636, 2019.

\bibitem{sirer2016online}
E.~G. Sirer.
\newblock Bitcoin guarantees strong, not eventual, consistency, 2016.
\newblock Available
  [\href{http://hackingdistributed.com/2016/03/01/bitcoin-guarantees-strong-not-eventual-consistency/}{online}].
  [Accessed: 19-2-2019].

\bibitem{Sm11}
H.~L. Smith and H.~R. Thieme.
\newblock {\em Dynamical systems and population persistence}, volume 118.
\newblock American Mathematical Soc., 2011.

\bibitem{walmart}
M.~Smith.
\newblock {In Wake of Romaine {E}.\ coli Scare, {W}almart Deploys Blockchain to
  Track Leafy Greens}.
\newblock CNN Business, 2018.
\newblock Available
  [\href{https://news.walmart.com/2018/09/24/in-wake-of-romaine-e-coli-scare-walmart-deploys-blockchain-to-track-leafy-greens}{online}].
  [Accessed: 25-2-2019].

\bibitem{So18}
Y.~Sompolinsky and A.~Zohar.
\newblock {Bitcoin's Underlying Incentives}.
\newblock {\em Commun. ACM}, 61(3):46--53, 2018.

\bibitem{St16}
M.~Steffel, E.~F. Williams, and J.~Perrmann-Graham.
\newblock {Passing the buck: Delegating choices to others to avoid
  responsibility and blame}.
\newblock {\em Organizational Behavior and Human Decision Processes},
  135:32--44, 2016.

\bibitem{St18}
N.~Stifter, A.~Judmayer, P.~Schindler, A.~Zamyatin, and E.~R. Weippl.
\newblock {Agreement with Satoshi -- On the Formalization of Nakamoto
  Consensus}.
\newblock {\em IACR Cryptology ePrint Archive}, 2018:400, 2018.

\bibitem{szalachowski2018towards}
P.~Szalachowski.
\newblock Towards more reliable {B}itcoin timestamps.
\newblock In {\em Crypto Valley Conference on Blockchain Technology (CVCBT)},
  2018.

\bibitem{szalachowski2019strongchain}
P.~Szalachowski, D.~Reijsbergen, I.~Homoliak, and S.~Sun.
\newblock {StrongChain}: Transparent and collaborative proof-of-work consensus.
\newblock In {\em 28th {USENIX} Security Symposium}, 2019.

\bibitem{rocket2018snowflake}
{Team Rocket}.
\newblock {Snowflake to {A}valanche: A novel metastable consensus protocol
  family for cryptocurrencies}, 2018.

\bibitem{Th18}
R.~Thurimella and Y.~Aahlad.
\newblock {The Hitchhiker's Guide to Blockchains: A Trust Based Taxonomy},
  2018.
\newblock Available
  [\href{https://wandisco.com/assets/whitepapers/the-hitchhikers-guide-to-blockchains.pdf}{online}].
  [Accessed: 15-11-2018].

\bibitem{US19online}
{U.S. Antitrust Division}.
\newblock Herfindahl-{H}irschman index.
\newblock Available
  [\href{https://www.justice.gov/atr/herfindahl-hirschman-index}{online}].
  [Accessed: 1-3-2019].

\bibitem{Va13}
V.~V. Vazirani.
\newblock {\em Approximation algorithms}.
\newblock Springer Science \& Business Media, 2013.

\bibitem{Med19}
T.~Vazz.
\newblock {The Real Blockchain Trilemma}, 2019.
\newblock Available
  [\href{https://medium.com/coinmonks/the-real-blockchain-trilemma-58824b52fe1d}{online}.
  [Accessed: 15-12-2019].

\bibitem{Ne44}
J.~von Neumann and O.~Morgenstern.
\newblock {\em Theory of Games and Economic Behavior}.
\newblock Princeton University Press, 1944.

\bibitem{Vu15}
M.~Vukoli{\'{c}}.
\newblock {The Quest for Scalable Blockchain Fabric: Proof-of-Work vs. BFT
  Replication}.
\newblock In J.~Camenisch and D.~Kesdo{\u{g}}an, editors, {\em Open Problems in
  Network Security}, pages 112--125. Springer International Publishing, 2016.

\bibitem{Wa13}
J.~R. Wallrabenstein and C.~Clifton.
\newblock {Equilibrium Concepts for Rational Multiparty Computation}.
\newblock In S.~K. Das, C.~Nita-Rotaru, and M.~Kantarcioglu, editors, {\em
  Decision and Game Theory for Security}, pages 226--245, Cham, 2013. Springer
  International Publishing.

\bibitem{Wa18}
W.~{Wang}, D.~T. {Hoang}, P.~{Hu}, Z.~{Xiong}, D.~{Niyato}, P.~{Wang}, and
  Y.~{Wen}.
\newblock {A Survey on Consensus Mechanisms and Mining Strategy Management in
  Blockchain Networks}.
\newblock {\em IEEE Access}, 2019.

\bibitem{Eth19}
E.~Wiki.
\newblock On sharding blockchains, 2019.
\newblock Availabel
  [\href{https://github.com/ethereum/wiki/wiki/Sharding-FAQ}{online}.
  [Accessed: 15-12-2019].

\bibitem{Wo89}
T.~Wong.
\newblock An application of game theory to corporate governance.
\newblock {\em Omega}, 17(1):59--67, 1989.

\bibitem{wust2018you}
K.~W{\"u}st and A.~Gervais.
\newblock Do you need a blockchain?
\newblock In {\em 2018 Crypto Valley Conference on Blockchain Technology
  (CVCBT)}, pages 45--54. IEEE, 2018.

\bibitem{Yl16}
J.~Yli-Huumo, D.~Ko, S.~Choi, S.~Park, and K.~Smolander.
\newblock {Where Is Current Research on Blockchain Technology? A Systematic
  Review.}
\newblock {\em PLoS ONE}, 11(10):e0163477, 2016.

\bibitem{Pz17}
V.~Zamfir.
\newblock personal communication.

\bibitem{Za18}
A.~Zamyatin, N.~Stifter, P.~Schindler, E.~R. Weippl, and W.~J. Knottenbelt.
\newblock {Flux: Revisiting Near Blocks for Proof-of-Work Blockchains}.
\newblock {\em IACR Cryptology ePrint Archive}, 2018:415, 2018.

\bibitem{Zh17}
R.~Zhang and B.~Preneel.
\newblock {Publish or Perish: A Backward-Compatible Defense Against Selfish
  Mining in Bitcoin}.
\newblock In {\em CT-RSA}, 2017.

\bibitem{Zh19}
R.~Zhang and B.~Preneel.
\newblock {Lay Down the Common Metrics: Evaluating Proof-of-Work Consensus
  Protocols' Security}.
\newblock In {\em Proceedings of the 2016 ACM SIGSAC Conference on Computer and
  Communications Security}, 2019.

\bibitem{Zh18}
Z.~Zheng, S.~Xie, X.~Chen, and H.~Wang.
\newblock {Blockchain Challenges and Opportunities: A Survey}.
\newblock {\em International Journal of Web and Grid Services (IJWGS)},
  14(4):352--375, 2018.

\bibitem{zilliqa}
{Zilliqa}.
\newblock Zilliqa.
\newblock [\href{https://zilliqa.com/}{website}]. [Accessed: 26-2-2019].

\bibitem{zilliqawp}
{Zilliqa Team}.
\newblock {The {Z}illiqa Technical Whitepaper}, 2017.
\newblock Available [\href{https://docs.zilliqa.com/whitepaper.pdf}{online}].
  [Accessed: 26-2-2019].

\bibitem{Zo17}
A.~Zohar.
\newblock {Securing and Scaling Cryptocurrencies}.
\newblock In {\em Proceedings of the Twenty-Sixth International Joint
  Conference on Artificial Intelligence, {IJCAI-17}}, pages 5161--5165, 2017.

\end{thebibliography}

\end{document}